\documentclass[article,superscriptaddress,amsmath,amssymb,aps,prx,floatfix,twocolumn]{revtex4-1}
\usepackage{graphicx}	
\usepackage{dcolumn}	
\usepackage{bm}
\usepackage{gensymb}	
\usepackage[dvipsnames]{xcolor}
\usepackage{csquotes} 

\usepackage{hyperref}	
\usepackage	[caption=false]	{subfig}
\usepackage	{graphicx, epstopdf, epsfig, rotating, floatpag, tikz}
\graphicspath{{figures/}}

\usepackage{ifthen}
\usepackage{color}
\usepackage{ulem}             
\definecolor{rot}{rgb}{1,0,0}
\definecolor{blau}{rgb}{0,0,1}
\definecolor{orange}{rgb}{.5,.5,0}
\definecolor{dunkelgruen}{rgb}{.133,0.545,0.133}

\newif\ifcom
\newif\ifdel
\comtrue               %
\delfalse

\begin{document}

\title{Origin and dynamics of umbrella states in rare-earth iron garnets}

\author{Bruno Tomasello}
\email{brunotomasello83@gmail.com}
\affiliation{Physics and Astronomy, Division of Natural Sciences, University of Kent, Canterbury, CT2 7NH, United Kingdom}
\author{Dan Mannix}
\email{dan.mannix@ess.eu}
\affiliation{European Spallation Source, SE-221 00 Lund, Sweden}
\affiliation{Institut N\'eel, CNRS Grenoble, Grenoble, 38042 France}
\author{Stephan Gepr\"ags}
\email{stephan.gepraegs@wmi.badw.de}
\affiliation{Walther-Mei{\ss}ner-Institut, Bayerische Akademie der Wissenschaften, 85748 Garching, Germany}
\author{Timothy Ziman}
\email{ziman@ill.fr}
\affiliation{Institut Laue Langevin, 38045 Grenoble Cedex 9,France}
\affiliation{Institute for Materials Research, Tohoku University, Sendai, Miyagi 980-8577, Japan}

\date{\today}

\begin{abstract}
	Rare-earth iron garnets $R_{3}$Fe$_{5}$O$_{12}$ are fascinating insulators with very diverse magnetic phases. Their strong potential in spintronic devices has encouraged a renewal of interest in the study of their low temperature spin structures and spin wave dynamics. A striking feature of rare-earth garnets with $R$-ions exhibiting strong crystal-field effects like Tb-, Dy-, Ho-, and Er-ions is the observation of low-temperature non-collinear magnetic structures featuring ``umbrella-like'' arrangements of the rare-earth magnetic moments. 
	In this study, we demonstrate that such umbrella magnetic states are naturally emerging from the crystal-field anisotropies of the rare-earth ions.
	By means of a general model endowed with only the necessary elements from the crystal structure, we show how umbrella-like spin structures can take place and calculate the canting-angle as a result of the competition between the exchange-interaction of the rare-earth and the iron ions as well as the crystal-field anisotropy. Our results are compared to experimental data, and a study of the polarised spin wave dynamics is presented. Our study improves the understanding of umbrella-like spin structures and paves the way for more complex spin wave calculations in rare-earth iron garnets with non-collinear magnetic phases. 
\end{abstract} 
 
\maketitle

\section{Introduction}
	Magnetic insulators in the family of rare-earth iron garnets $R_{3}$Fe$_{5}$O$_{12}$ (where $R$ is a trivalent rare-earth ion) are currently seeing a renewal of interest, in large part because of their interesting physical properties and possible applications in the fields of spintronics, spin caloritronics and magnonics~\cite{Kruglyak:2010,Bauer:2012,Han:2018,Brataas:2020}. In particular, Yttrium Iron Garnet (Y$_{3}$Fe$_{5}$O$_{12}$, YIG) has a ferrimagnetic structure stable well above room temperatures -- with a Curie temperature of 560 K -- and superior microwave properties with extremely long lifetime and diffusion length of magnons~\cite{Cornelissen:2015,Jamison:2019}. It is therefore a material of choice for microwave devices and in the field of magnonics~\cite{Serga:2010}. However, there is much interest in looking at other members of the same $R_{3}$Fe$_{5}$O$_{12}$-family by substituting the non-magnetic Y$^{3+}$-ions in YIG with other rare-earth ions: $R$ = Gd, Tb, Dy, Ho, Er, Tm, Yb or Lu. These rare-earth ions now bear their own magnetic moments, which, in contrast to the iron moments, have strong spin-orbit coupling and crystal-field splittings~\cite{Newman:1969}. This may lead to a variety of magnetic phases with some of the attractive features of YIG but differences which may be interesting for specific applications. This was recognised soon after the magnetic structures were elucidated by N{\'e}el, Bertaut and collaborators in the mid 1950's, who commented on the fact that large single crystals with very narrow resonance line-widths were common to all rare-earth iron garnets and favoured applications in magneto-optics~\cite{Neel:1964}. They also found that while the rare-earth ion did not alter the temperature of ferrimagnetic order very much -- it remains  between 548\,K and 578\,K for $R$= Gd, Tb, Dy, Ho, Er, and Tm  -- there was a lower ``compensation" temperature where the remanent magnetization of the rare-earth ion exactly cancel the net iron magnetization~\cite{Pauthenet:1958}. N\'eel suggested very soon that this implied a magnetic model where the rare-earth moment had an exchange with the iron moments that was weaker than the exchanges within the iron sites leading to ferrimagnetism~\cite{Neel:1954}. The rare-earth ion can therefore be treated as an ``exchange-enhanced'' paramagnetic moment with a strong temperature dependence~\cite{Belov:1996}.

	The magnetic structure of rare-earth iron garnets $R_{3}$Fe$_{5}$O$_{12}$ is a consequence of the antiferromagnetic super-exchange between the iron moments on the tetrahedrally-coordinated $d$-sites and that of the octahedrally-coordinated $a$-sites as well as the rare-earth moment on the dodecahedrally-coordinated $c$-sites of the garnet unit cell~\cite{Geller:1957a}. While the magnetic structure of $R_{3}$Fe$_{5}$O$_{12}$ has been described in great detail, primarily by neutron diffraction~\cite{Tcheou:1970,Pickart:1970,Guillot:1982,Lahoubi:1984,Hock:1990}, only recently the spin wave dynamics came into focus of research~\cite{Princep:2013,Barker:2016,Shen:2019,Shamoto:2020}. For example, the chirality of the spin waves in YIG, while long expected at least at zero temperature, was only recently explicitly demonstrated by polarisation analysis of inelastic neutron scattering, at both low and room temperatures~\cite{Nambu:2020}. 
	The observed opposite chirality, contrasting the acoustic and the optical spin-wave modes, is expected to have great influence on experiments in the fields of spintronics and magnonics.
	This was demonstrated in Gd$_{3}$Fe$_{5}$O$_{12}$ that shows  an additional sign change of the spin Seebeck voltage at low temperatures~\cite{Geprags:2016}. The influence of the rare-earth moments on the chiral properties of the spin waves in $R_{3}$Fe$_{5}$O$_{12}$ is therefore of paramount importance. 

	In addition to the  collinear magnetic structure between the rare-earth and iron moments we have just described, an ``umbrella"-like structure is observed at low temperatures in rare-earth iron garnets $R_{3}$Fe$_{5}$O$_{12}$ with $R$-ions exhibiting strong crystal-field effects
	~\cite{Pickart:1970,Guillot:1982,Hock:1991,Lahoubi:2012}. Such a non-collinear structure is  based on canting of the $R$-moments with respect to the iron-moments at some angle $\theta_R$. In this paper, we introduce a simple model Hamiltonian to describe this non-collinear magnetic structure and to explore the consequences on the spin wave dynamics. The ``umbrella"-like structure are seen to arise quite naturally from the competition of the strong single-ion anisotropies of different rare-earth ions and the exchange interaction between the rare-earth and iron moments.

\section{Simple model for Umbrella States}
\label{sec:simplemodel}

\begin{figure}[tb]
\centering
\includegraphics[width=\linewidth]{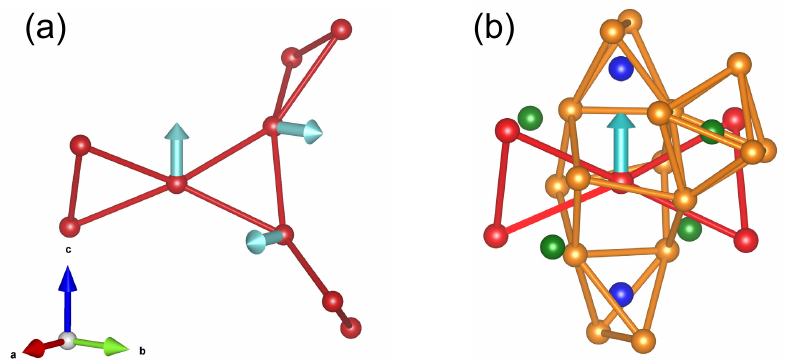} \hspace{4mm}
\caption{{\bf(a)} Portion of the hyperkagome lattice (red spheres and links), with single-ion anisotropies (light blue arrows) shown for three next-nearest neighbour rare-earth $c$-sites. The easy-axes, each parallel to a given crystallographic direction, are obtained by exact diagonalisation of the $D_{2}$ crystal-field Hamiltonians, for example from Tb$_{3}$Fe$_{5}$O$_{12}$ in Ref.~\cite{Hau:1986}.
	{\bf(b)} Local view of the complex crystal structure of rare-earth iron garnets. We focus on one rare-earth ion of interest (red-sphere with light blue easy-axis) and its neighbouring atoms. The neighbouring rare-earth $c$-sites with its links are shown in red, as in the left panel. The dark-yellow spheres are the oxygen-ions ($h$-sites), the next nearest neighbours to the central $R$ site form a distorted cube and they have a dodecahedral point-group symmetry ($D_2$). The blue spheres are the two Fe $d$-sites nearest to the $c$-site of interest, and are shown in their tetrahedral oxygen environment. The green spheres are the Fe $a$-sites nearest to the same central $c$-site, and for simplicity only one is shown in its octahedral oxygen environment.}
\label{fig:AnisotropiesTriangle}
\end{figure}

	To investigate the influence of the non-collinear ``umbrella"-state on the spin wave dynamics in rare-earth iron garnets $R_{3}$Fe$_{5}$O$_{12}$, we introduce a simple model Hamiltonian in the following, by taking into account single-ion anisotropies of rare earth $R^{3+}$-ions.	From the structure of the garnet unit cell, described by the space group $Ia\overline{3}d$, where the $R^{3+}$-ions are dodecahedrally-coordinated ($c$-sites) and the Fe$^{3+}$-ions tetrahedrally- ($d$-sites) as well as octahedrally-coordinated ($a$-sites), the easy-axes of the rare-earth ions automatically lie in three orthogonal directions within the unit cell (see Fig.\ref{fig:AnisotropiesTriangle}(a)). When we consider the coupling between the $R^{3+}$- and the neighbouring Fe$^{3+}$-moments of the majority $d$-sublattice (blue spheres in Fig.\ref{fig:AnisotropiesTriangle}(b)), these act as torque and the $R^{3+}$ moments tilt away from the crystalline anisotropy, and in favour of the $[111]$ molecular field from the Fe$^{3+}$-$d$-site. While the existence of the three orthogonal two-fold axes is a rigorous consequence of the $D_2$ symmetry, the selection of one of these as the easy axis can be seen intuitively from the positioning of the oxygen O$^{2-}$-ions with respect to the nearest Fe$^{3+}$ $a$-sites (see Fig.\ref{fig:AnisotropiesTriangle}(b)). For example, the ground state of the crystal-field of Tb$_{3}$Fe$_{5}$O$_{12}$ has the easy axis (light blue arrows in Fig.\ref{fig:AnisotropiesTriangle}) aligned along the direction connecting a Tb$^{3+}$-$c$-site with its neighbouring Fe$^{3+}$-$d$-sites~\cite{Hau:1986,Wawrzynczak:2019}.
As the $c$-$d$ distances (3.11~\AA) are comparable to the $c$-$h$ distances (of the eight $h$ O$^{2-}$-ions in proximity of a $c$-site, four sit at 2.4~\AA, and four at 2.5~\AA) the electrostatic charges of the Fe$^{3+}$ ions on the $d$-sites might influence the orientation of the local easy axis.

\begin{figure}[tb]
  \centering
  \includegraphics[width=\linewidth]{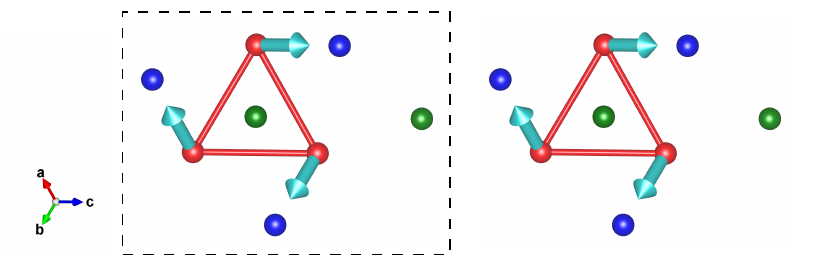}
  \caption{The one-dimensional 8-sublattice model. The dashed line shows the unit cell. In accordance with Fig.\ref{fig:AnisotropiesTriangle}, rare-earth ions (red spheres) are shown with magnetic anisotropies (light-blue arrows) calculated from crystal-field Hamiltonians~\cite{Hau:1986,Wawrzynczak:2019}. The  iron ions (blue $d$-sites, green $a$-sites) have magnetic moments (not shown) along the [111] direction according to their ferrimagnetic ordered phase. 
  }
  \label{fig:8sublattice}
\end{figure}

	In order to understand the effects of the coupling on the dynamics, including the frequencies and  chirality of spin wave propagation at finite wave vectors, we introduce dispersion by coupling the moments in a one-dimensional form. The model can be considered as a minimal extension of that by Tinkham~\cite{{Tinkham:1961}}. Tinkham considered the coupled dynamics of rare-earth iron garnets at low frequencies, where he reduced the system to two effective moments, one representing the rare-earth moment and the other the net iron magnetic moment. The crucial element to allow for the non-collinear umbrella structure is to include at least three rare-earth dodecahedral $c$-sites with the three orthogonal directions and the corresponding number of Fe-$d$ and Fe-$a$-sites to respect stochiometry. Therefore, the minimum cell with three different anisotropy directions for the $R^{3+}$ $c$-site, and compatible with the stochiometry of the magnetic sites in $R_{3}$Fe$_{5}$O$_{12}$, would have a total of 8 magnetic sites: 3 $R^{3+}$-moments with spin $\mathbf{S}$, 3 tetrahedral Fe$^{3+}$-$d$ moments with spin $\mathbf{s}_{d}$, and 2 octahedral Fe$^{3+}$-$a$ moments with spin $\mathbf{s}_{a}$. We consider a one-dimensional chain based on such 8-sublattice unit cell, as represented in Fig.~\ref{fig:8sublattice}. As in the real compounds, the different numbers of $d$- and $a$-sites give ferrimagnetism quite naturally.   

	Our simplest Hamiltonian reads
\begin{equation}
\begin{split}
{\cal H}^{0} =   
	- & K_e \sum_{i} ({S}_{i}^{\alpha_i})^2 \\
	  & +J_{cd}\sum_{\langle ij\rangle} {\bf S}_{i} \cdot{\bf s}^{d}_{ j} +J_{ad}\sum_{\langle j k \rangle} {\bf s}^{d}_{j} \cdot{\bf s}^{a}_{k}\ .
\end{split}
\label{eq:HDKJ0}
\end{equation}
where $K_e>0$ regulates the easy-axis anisotropy of the rare-earth $c$- moments, while $J_{ad}$ and $J_{cd}$ are exchange constants.
The model is defined by these three couplings and of course also by the lattice of sites considered. The summation $\langle ij\rangle$ runs over pairs of rare-earth and iron $d$ moments considered as nearest-neighbours, and ${\langle j k \rangle}$ over nearest neighbour tetrahedral-octahedral pairs, again defined by whichever model lattice we use. The strength of the magnetic moments correspond to the total angular momenta of the free ions (see further discussion in Section~\ref{sec:comparisonrealmaterials}), hence $|\mathbf{s}_{a}|=|\mathbf{s}_{d}|= 5/2$, and $|\mathbf{S}|=$ the total angular momentum of the $R^{3+}$ of interest.
In equation~\eqref{eq:HDKJ0}, there are three possible values of the vector component $\alpha_i = x, y$ or $z$ for the three next-nearest neighbour rare-earth sites at the vertices of the same triangle. These vector components, as shown in Figs.~\ref{fig:AnisotropiesTriangle},\ref{fig:8sublattice}, define three mutually perpendicular easy-axis anisotropies on each triangle. 
	In the model Hamiltonian ${\cal H}^{0}$, we consider only nearest-neighbor interactions and, in particular, neglect exchange interaction between the rare-earth ions $J_{cc}$, and the rare-earth ion and the iron ${a}$-site $J_{ca}$. 

	While the model defined by equation~\eqref{eq:HDKJ0} is the simplest to look at, further terms ought to be included to investigate quantitative features of the magnetic structure and the spin dynamics. These include exchange interactions $J_{aa}$ and $J_{dd}$ within the iron ${a}$- and ${d}$-sites, respectively, but also hard-axes terms with anisotropy constant $K_h$ accounting for the complex constraints from the crystalline fields. These are gathered in the following Hamiltonian
\begin{equation}
\begin{split}
	{\cal H}^{1} = \;
	& K_h \sum_{i} ({S}_{i}^{\beta_i})^2 \\
	& + J_{dd} \sum_{\langle j j' \rangle} {\bf s}_{j}^d\cdot{\bf s}_{j'} ^d
	 + J_{aa} \sum_{\langle k k' \rangle} {\bf s}_{k}^a\cdot{\bf s}_{k'}^a,
\end{split}
\label{eq:HDKJ1}
\end{equation}
which includes a magnetically hard axis $\beta_i$ orthogonal to the easy axis $\alpha_i$ of the $i$-th rare-earth site ($K_{h}>0$, if $K_e$ is taken to be positive in ${\cal H}^{0}$). To simplify the numerical calculations we can take  $\beta_i$ also orthogonal to the Fe moments so that it does not influence the canting. The effect of the additional magnetic anisotropy $K_{h}$ will not be discussed further in this paper,  except in   the discussion of ``double-umbrella" spin structures found in Tb$_{3}$Fe$_{5}$O$_{12}$ and Yb$_{3}$Fe$_{5}$O$_{12}$ (see Section~\ref{sec:hardaxisdoubleumbrella}). It will also influence the chirality of excitations, but this will be left to a future, more detailed, publication.

	If we fix the spin values of the rare-earth and iron moments and first neglect $K_h$, there are five parameters left for the minimal Hamiltonian ${\cal H}= {\cal H}^{0} + {\cal H}^{1}$. These parameters can be set initially by different observables of a compound showing an umbrella structure:  
\begin{enumerate}
	\item The ordering temperature for ferrimagnetism or alternatively the dispersion of the lowest energy mode in $R_{3}$Fe$_{5}$O$_{12}$: this constrains estimates of $J_{ad}$, $J_{dd}$, and $J_{aa}$ for a given lattice. Since the Curie temperature $T_{\rm C}$ is almost independent of the rare-earth $R^{3+}$-ion, a very good first approximation is to take these values from YIG. Recently, $J_{ad}$, $J_{dd}$, and $J_{aa}$ have been estimated from the spin-wave dispersion of YIG measured by inelastic neutron scattering~\cite{Princep:2017,Nambu:2020}. We therefore take 6.8\,meV, 0.52\,meV, and 1.1\,meV for $J_{ad}$, $J_{dd}$, and $J_{aa}$, respectively.
	\item The gap $\Delta_{\rm CF}$ between the ground state doublet (or quasi-doublet for non-Kramers systems) and the first excited state of the crystal-field spectrum: this determines $K_e$~\cite{Hau:1986,Wawrzynczak:2019} for each compound.
	\item The magnetic compensation temperature, where the total remanent magnetisation of the garnet compound vanishes: this gives an estimate of the ratio $J_{cd}/J_{ad}$. We should emphasize that $J_{cd}$ can vary depending on the rare-earth considered (see Section \ref{sec:comparisonrealmaterials}), however, in general $|J_{cd}/J_{ad}|<1$.
\end{enumerate} 

	In the remaining of this paper the use of superscript or subscript for the kind of spins is interchanged for notational convenience (${\bf s}^{d}\equiv{\bf s}_{d}$), and the $c$ label for the rare-earth site is omitted (${\bf S}\equiv{\bf S}_{c}$), as in Eqs.~\eqref{eq:HDKJ0}, \eqref{eq:HDKJ1}.

\section{Ground state properties}
\label{sec:groundstate}

	In our one-dimensional spin model each iron Fe$^{3+}$-$d$ moment is coupled to a single rare-earth $R^{3+}$-$c$ moment canted in one of three directions ($\alpha_i = x, y$ or $z$). This will induce an additional canting $\theta_\mathrm{Fe}$ of the Fe$^{3+}$-moments, {\it i.e.}, in this model the Fe moments themselves have a small ``umbrella-like'' spin structure induced by the rare-earth $R^{3+}$-moments. However, the canting angle $\theta_\mathrm{Fe}$ is expected to be small since the ratio $J_{cd}/J_{ad}$ is known to be small. Moreover, the exchange interaction within the iron $d$-sites and $a$-sites ($J_{dd}$ and $J_{aa}$, respectively) further reduces this secondary canting. 

	As the single-ion anisotropy of the rare-earth ions pull towards orthogonal directions, the net magnetic moment remains at the [111]-directions of the garnet structure and we can minimize the pairs of terms of ${\cal H}= {\cal H}^{0} + {\cal H}^{1}$ separately with respect to the canting angle of the rare-earth $R^{3+}$-moments $\theta_R$ and the secondary canting of the Fe$^{3+}$-moments $\theta_\mathrm{Fe}$. If we define $\alpha_o$ as the angle of the three orthogonal axes of the net moment with respect to the [111]-direction, $\alpha_o = \cos^{-1}  (1/\sqrt{3})$, then Eq.~\eqref{eq:HDKJ1} yields the classical energy as a function of the canting angles $\theta_R$ and $\theta_\mathrm{Fe}$
\begin{equation}
\begin{split}
\epsilon (\theta_R,\theta_\mathrm{Fe})=  & -K_e S^2\cos^2(\alpha_o - \theta_R)  + J_{cd} S s_{d}\cos(\theta_R-\theta_\mathrm{Fe}) \\
&   +J_{dd}s_{d}^2(2\cos^2 \theta_\mathrm{Fe} - \sin^2 \theta_\mathrm{Fe})/2  \\
&   -2 J_{ad} \; s_{d} s_{a}\cos^2 \theta_\mathrm{Fe} \ .
\end{split} 
\label{eq:sublatticeenergy8}
\end{equation}
 
\begin{figure}[tb]
\centering
\includegraphics[width=1.0\linewidth]{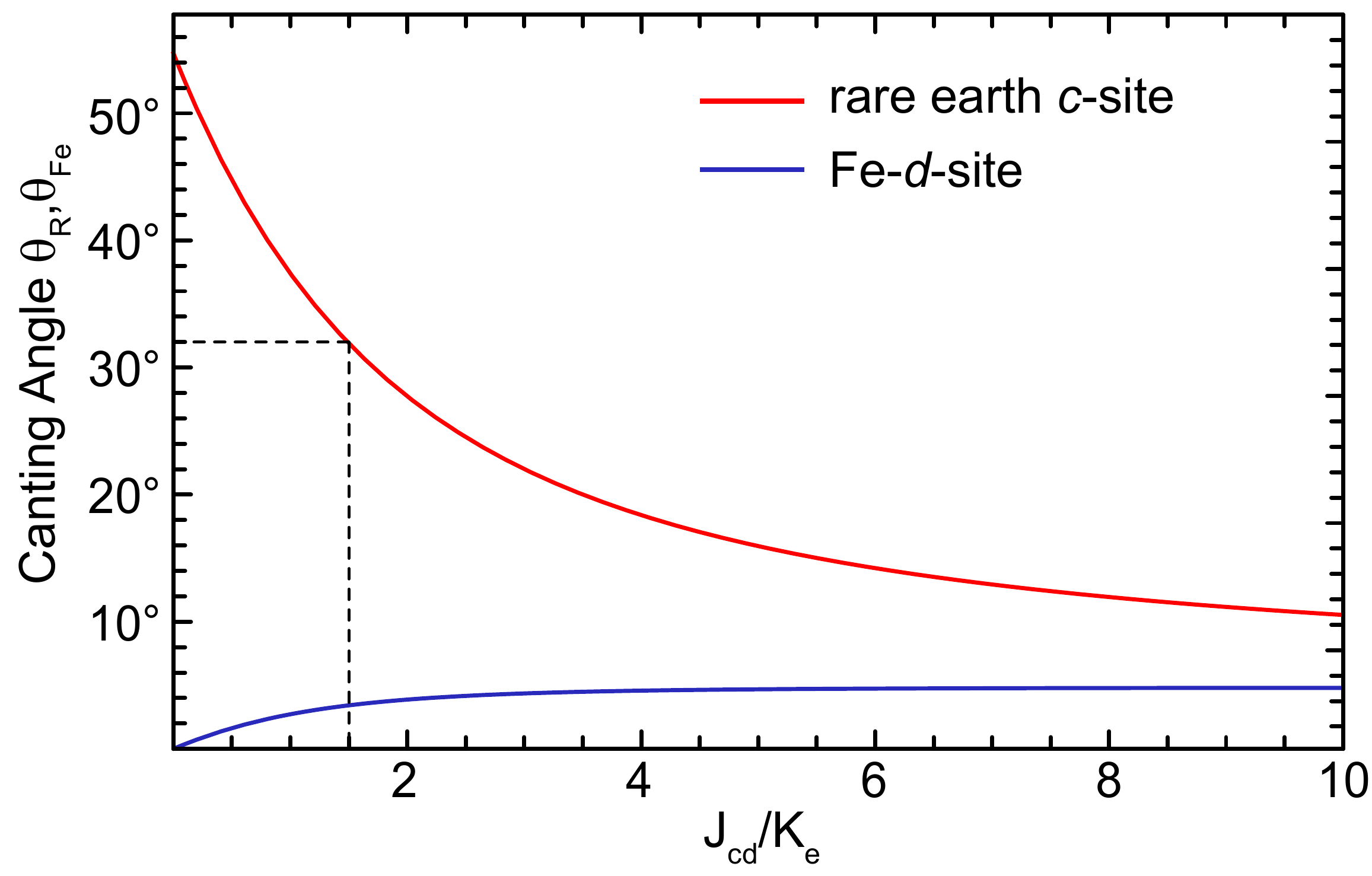} \hspace{4mm}
\caption{Canting angles $\theta_R$ (red line) and $\theta_\mathrm{Fe}$ (blue line) of the rare-earth $c$-moments and the iron $d$-moments, respectively, as a function of the dimensionless ratio of $c$-$d$ exchange $J_{cd}$ to single-ion anisotropy $K_e$ calculated using the spin values $|\mathbf{s}_{a}|=|\mathbf{s}_{d}|= 5/2$ and $|\mathbf{S}|=6$. Due to the interaction with the rare-earth moments, the Fe-$d$ moments also become slightly non-collinear. The dashed lines indicate a canting angle $\theta_R=32^\circ$, which is used for the calculation of the spin-wave dynamics in Section~\ref{sec:spinwaves}.}
\label{fig:canting}
\end{figure}

	Minimizing the energy $\epsilon (\theta_R,\theta_\mathrm{Fe})$ gives $\theta_R$ and $\theta_\mathrm{Fe}$ as a function of $J_{cd}s_{d}/\left( K_e S\right)$. The thus obtained canting angles $\theta_R$ and $\theta_\mathrm{Fe}$ are shown in Fig.~\ref{fig:canting} as a function of the dimensionless ratio of the $c$-$d$-exchange $J_{cd}$ to the single-ion anisotropy $K_e$. A large canting of the rare-earth moments is expected for a large single-ion anisotropy and weak $c$-$d$ coupling. By increasing the exchange interaction $J_{cd}$ the Fe-$d$-moments also become slightly non-collinear. However, we note that we do not consider any crystalline anisotropy on the Fe-sites, since this is expected to be weak~\cite{Cherepanov:1993}.    

\subsection*{Linear Spin wave dynamics}
\label{sec:spinwaves}

\begin{figure}[tb]
\centering
\includegraphics[width=0.88\linewidth]{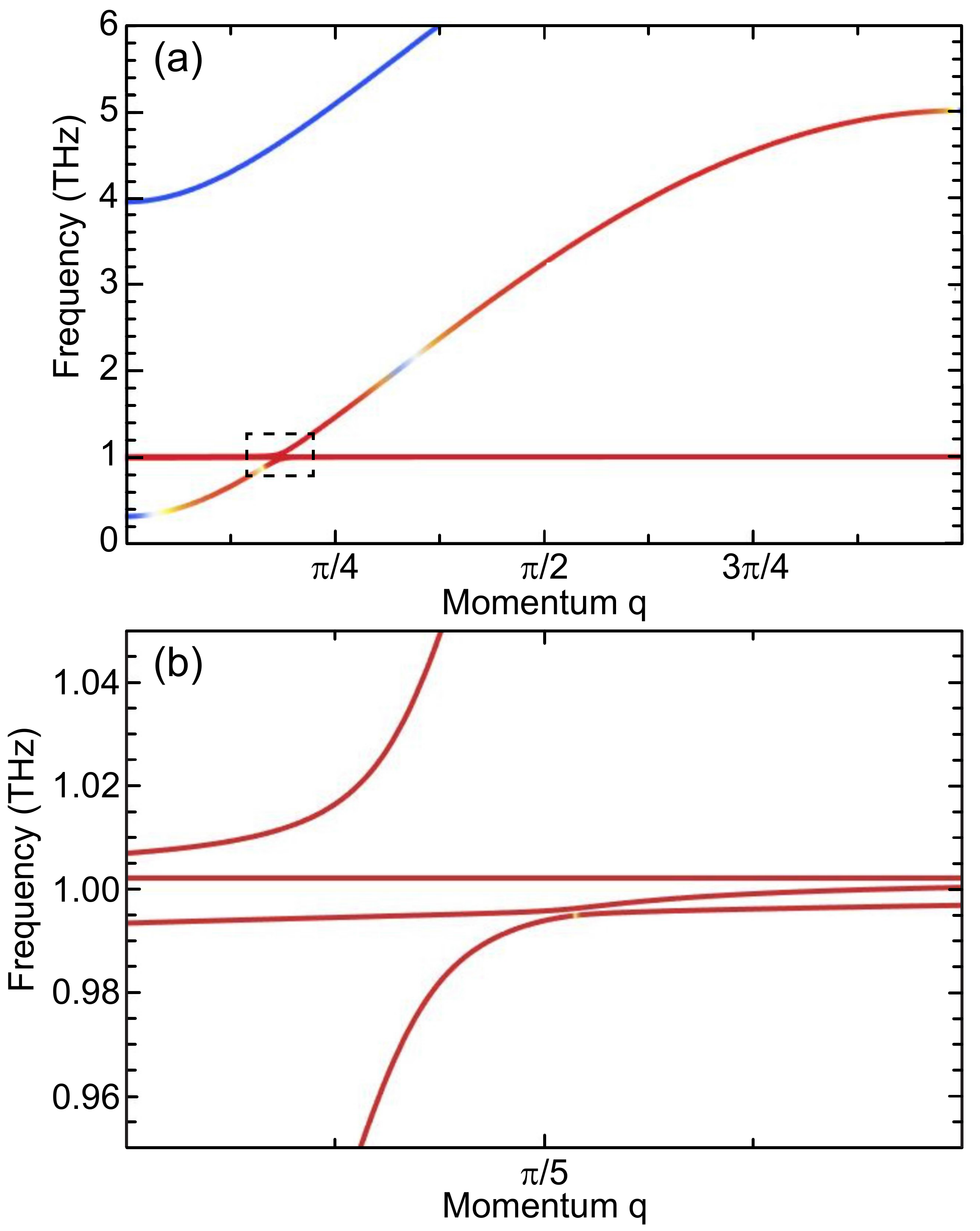} \hspace{0mm}
\caption{(a), (b) Spin wave dispersion and chirality calculated using Eq.~\eqref{eq:HDKJ0} assuming a weak $c$-$d$-coupling with $J_{cd} = 0.05 J_{ad}$. The gap $\Delta_{\rm CF}$ between the ground state doublet and the first excited state of the crystal-field spectrum of the $R^{3+}$-ions is assumed to be 4.2\,meV corresponding to 1\,THz. The chirality $\mathcal{X}$ is colour coded with the extreme values $\mathcal{X}=-1$ (blue) and $\mathcal{X}=+1$ (red). The momentum vector $\mathbf{q}$ is in units of the reciprocal lattice of the unit cell. (b) Enlargement of the region of the avoided crossing indicated by the dashed box in (a).}
\label{fig:spinwaves1}
\end{figure}

	Once we have determined the ground state of our 8-sublattice model in a mean field approach, we now can calculate the dynamical structure factor via linear spin wave theory~\cite{Cooper:1962}. Because of the interest in the chirality, especially of the low energy spin waves, we calculate the spin polarized cross-sections with  the sign of chirality defined with respect to the spin axis  along  the [111] direction, {\rm i.e.} the direction of the Fe-$a$-moments. To introduce the main qualitative features, we will restrict our discussion on the results calculated using the basic Hamiltonian introduced in Eq.~\eqref{eq:HDKJ0}. In Figure~\ref{fig:spinwaves1}(a) we show the dispersion of the lowest five modes of our simple one-dimensional model for a small $c$-$d$ coupling with $J_{cd} = 0.05 J_{ad}$. This implies a relatively weak coupling between the acoustic as well as optic modes on the Fe-ions and the essentially local modes on the rare-earth $R^{3+}$-ions. The gap $\Delta_{\rm CF}$ between the ground state doublet and the first excited state of the crystal-field spectrum of the $R^{3+}$-ions is assumed to be 4.2\,meV. As there are three rare-earth ions per unit cell, the weakly dispersive single-ion mode at around 1\,THz is 3-fold degenerate. Figure~\ref{fig:spinwaves1}(b) shows an enlargement of the region around the avoided crossing of the original acoustic branch and the single-ion levels, demonstrating that in fact there is level repulsion between the upper and lower levels, while the intermediate level is apparently not hybridized. The property of this intermediate level is a consequence of a reflection symmetry of the one-dimensional model. Hybridization of the acoustic mode with the rare-earth ions,  which are strongly anisotropic, leads to a gap in the acoustic mode - there is no longer a Goldstone mode as rotational symmetry is broken. In the two figures the dispersion is colour-coded in terms of the chirality $\mathcal{X}$ of the modes, defined simply as
\begin{equation}
\mathcal{X} ({\bf q},\omega ) = \frac{\chi^{+-}({\bf q},\omega )-\chi^{-+}({\bf q},\omega )}{\chi^{+-}({\bf q},\omega )+\chi^{-+}({\bf q},\omega )} \, ,
\end{equation} 
where $\chi^{-+}$ and $\chi^{+-}$ are, respectively, the spin-flip {\it up} and spin-flip {\it down} dynamical structure factors. Therefore, the upper, optical mode has a chirality $\mathcal{X}=-1$ (full blue) and so only the $\chi^{-+}$ cross-section contributes. Instead, the lower, acoustic mode, which would have a chirality $\mathcal{X}=+1$ (full red) in the decoupled limit ($J_{cd}=0$), now develops a more complex, $q$-dependent, chirality $\mathcal{X} ({\bf q},\omega)$ due to its hybridization with the rare-earth ions.

\begin{figure}[tb]
\centering
\includegraphics[width=0.88\linewidth]{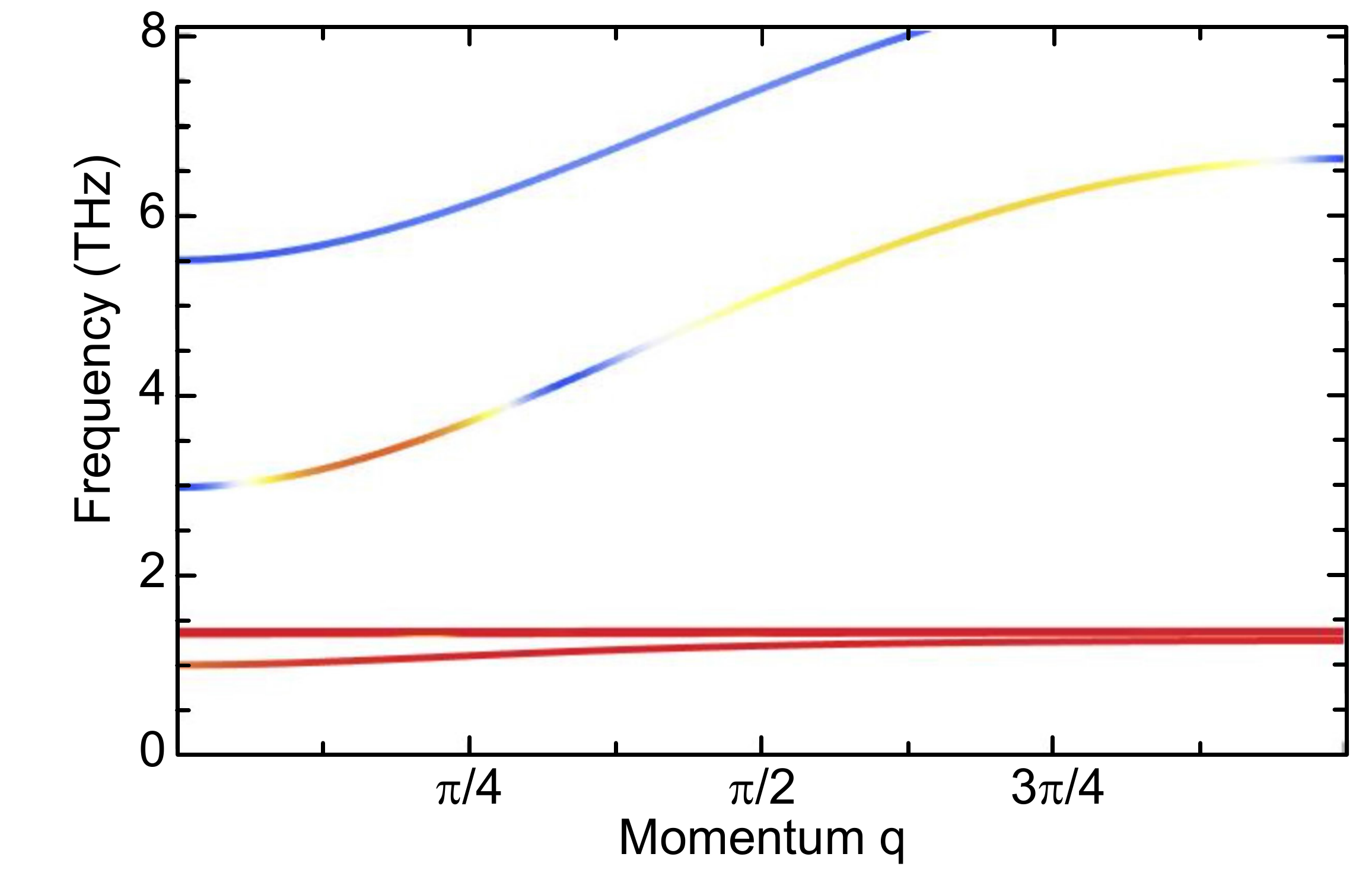} \hspace{0mm}
\caption{Spin wave dispersion and chirality (colour coded) calculated using Eq.~\eqref{eq:HDKJ0} assuming a stronger $c$-$d$-coupling $J_{cd} = 0.2 J_{ad}$, which is a reasonable value~\cite{Harris:1963}. The originally gapless fundamental Goldstone mode is pushed above the single-ion levels of the $R^{3+}$-ions, but still exhibiting a complex chirality $\mathcal{X} ({\bf q},\omega)$.} 
\label{fig:spinwaves2}
\end{figure}

	If we now increase the coupling strength $J_{cd}$ to a value that gives an umbrella structure with a canting angle of $\theta_R=32^\circ$ of the rare-earth ions (see dashed lines in Fig.~\ref{fig:canting}) and $J_{cd}=0.2J_{ad}$, which is a reasonable value~\cite{Harris:1963}, the originally gapless Goldstone mode is actually pushed above the single-ion levels, but retains a complex chirality $\mathcal{X} ({\bf q},\omega)$ (see Fig.~\ref{fig:spinwaves2}). This may be compared to Ref.~\onlinecite{Ohnuma:2013} where there is no-anisotropy  and a collinear structure but level repulsion gives a gapped ``acoustic'' mode.

\section{Comparison to  real materials}
\label{sec:comparisonrealmaterials}

	While the one-dimensional spin lattice is obviously simplified structurally compared to real $R_{3}$Fe$_{5}$O$_{12}$ materials, there are further simplifications that are inherent in the formulation. The rare-earth ``spin'' $S$ should be identified as the total angular momentum $J=L+S$ where the orbital $L$ and spin $S$ are strongly coupled by $L\cdot S$ coupling. By writing a single quadratic term in Eqs.~\eqref{eq:HDKJ0} and \eqref{eq:HDKJ1} we are implying that the single-ion ground state is a doublet. This is exact only for half-integer Kramers ions, naturally having a crystal-field spectrum of doublets, whereas for integer angular momenta (non-Kramers ions) the ground state magnetic moment originates from a crystal-field quasi-doublet of two low-lying singlets. In practice the aim is to benefit from an effective model, Eq.~\eqref{eq:HDKJ0}, accounting for the physics of a single-ion ground-state doublet; this is valid also for non-Kramers compounds (e.g. Tb$_{3}$Fe$_{5}$O$_{12}$) in the regime where the coupling $J_{cd}$ is sufficient to induce magnetic order thanks to the mixing of the low-lying singlets of the quasi-doublet~\cite{Wawrzynczak:2019}. In fact, each rare-earth compound has crystal-field parameters multiplying the general Steven's operators up to the order determined by the total angular momentum. As we treat the Hamiltonian in {\it linear} spin wave theory, this would renormalise some of the diagonal terms in the matrices without changing their structure substantially, but would clearly distort predictions of higher spin wave branches.

	We include  the anisotropy  only via the single-ion term and take {\it isotropic exchange} for the interaction between ions. This may be accurate for the Fe-Fe exchange ($J_{ad}$, $J_{aa}$, and $J_{dd}$), where spin-orbit is relatively weak, but is clearly an approximation for the rare-earth iron exchange $J_{cd}$ and is retained for simplicity only. To lowest order, the super-exchange $J_{\rm super}$ between the rare-earth ions and the Fe will involve only the total $S_i$ component of the $R^{3+}$ ion, and thus our parameter $J_{cd}$ should be understood as $(g_{J}-1)*J_{\rm super}$ with the Land\'e $g$-factor of the $R^{3+}$ considered. 
	We note the very recent study of Ref.~\cite{Pecanha-Antonio:2022}, where fitting to inelastic neutron scattering on Yb$_{3}$Fe$_{5}$O$_{12}$ was optimised including also {\it anisotropic exchange}. This is in accordance with earlier optical spectroscopy for this particular compound by Wickersheim and White~\cite{Wickersheim:1960,Wickersheim:1961}. It remains to be seen for other compounds whether the anisotropy in the exchange is as crucial as the full crystal field parameters.

	Notwithstanding the simplifications noted, the results shown in Figs.~\ref{fig:spinwaves1} and ~\ref{fig:spinwaves2} point to new and interesting features to be expected in the dynamics of umbrella structures compared to those of YIG and GdIG. The non-collinearity means that even at low temperatures we can no longer expect individual branches to have constant chirality, as has been demonstrated in YIG~\cite{Nambu:2020} and would be expected in the collinear GdIG~\cite{Shen:2019}. The effect seems to be strongest in the acoustic mode which now has strongly changing $q$-dependent chirality. As seen in Ref.~\cite{Nambu:2020} a still modest coupling between the rare-earth and Fe-d is enough to push the acoustic mode above the single-ion levels. We emphasise that, since for the dispersion we only used the Hamiltonian ${\cal H}^{0}$ in Eq.~\eqref{eq:HDKJ0}, the exact quantitative values at which this happens should not be considered precise for real materials (where $J_{aa}$ and $J_{dd}$ are non-zero).
	
\subsection*{Magnetic moment at finite temperature}
\label{sec:magneticmomentreal}

\begin{figure}[tb]
\centering
\includegraphics[width=.88\linewidth]{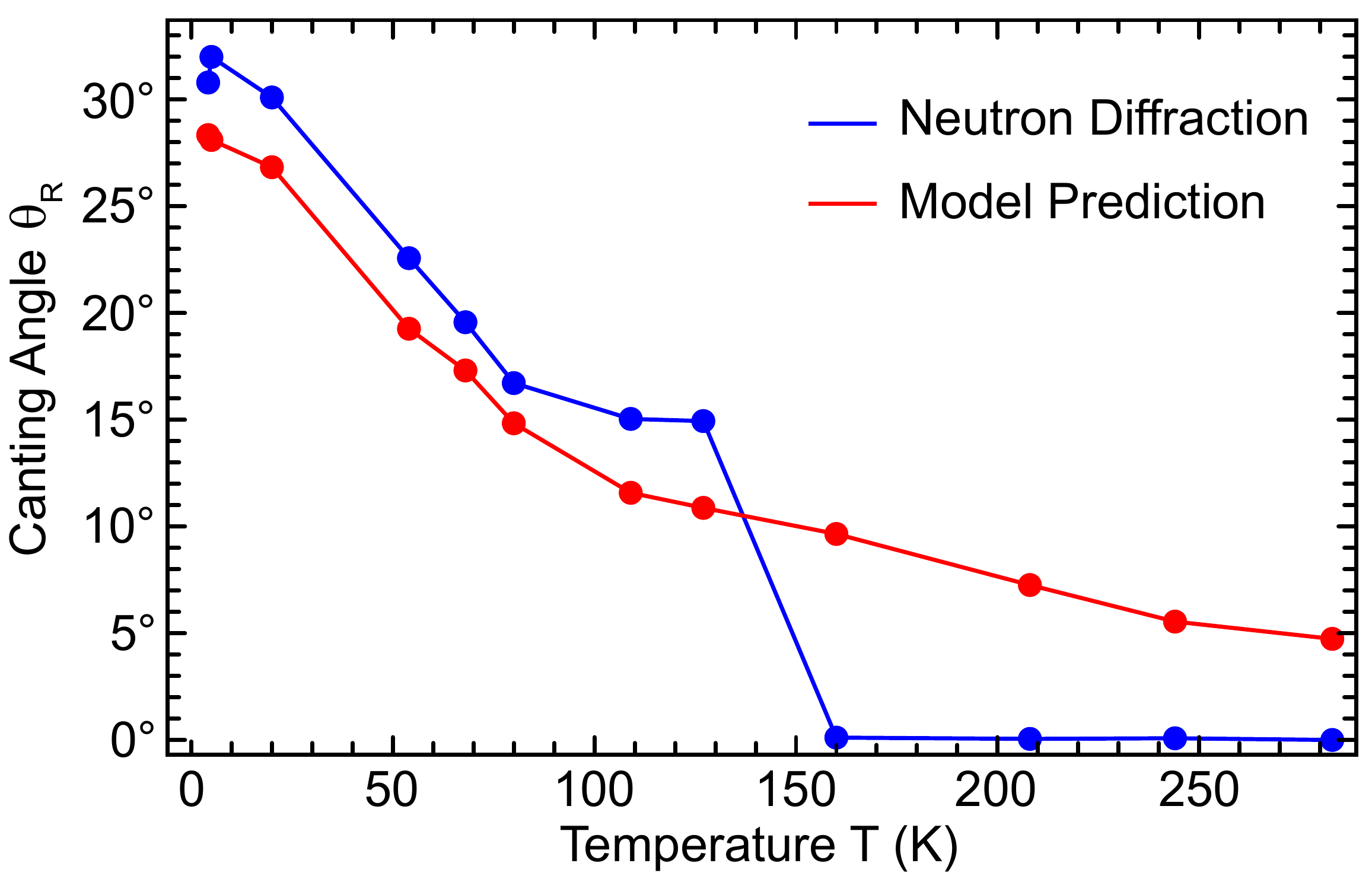} \hspace{0mm}
\caption{Mean field theory of the canting angle $\theta_R$ of Tb$_{3}$Fe$_{5}$O$_{12}$ (red symbols) calculated based on measured magnetization data from Ref.~\cite{Lahoubi:2015}. Contrary to the canting angle $\theta_R$ derived from Neutron diffraction measurement (blue symbols) no transition between a non-collinear ``umbrella"-like spin state to a collinear arrangement at 150\,K is expected.}
\label{fig:meanfieldcanting_TbIG}
\end{figure}

	To have a very simple treatment of finite temperature effects we will simply include a renormalization of the effective rare-earth moment, treated in the molecular field of the more strongly coupled Fe moments. Therefore, within a  mean field theory, we can estimate the canting angle of the rare-earth moments $\theta_R$ as a function of temperature by simply substituting $S$ by an effective magnetic moment $m_R(T)$ and take temperature independent estimates of $K_e$. The temperature dependent rare-earth moment $m_R(T)$ can be obtained by neutron diffraction or SQUID magnetometry measurements~\cite{Lahoubi:2012b,Pauthenet:1958}. For Tb$_{3}$Fe$_{5}$O$_{12}$, for example, the temperature evolution of $\theta_R$ calculated, again using  $J_{cd}= 0.2J_{ad}$, can be compared to $\theta_R$ from one of the umbrella by neutron diffraction experiments~\cite{Lahoubi:2012b}. As shown in Fig.~\ref{fig:meanfieldcanting_TbIG}, given such a simplistic approach, the agreement is remarkable. There is, however, no prediction of a transition to a collinear state above 150\,K, contrary to the past experimental interpretation, which is based on a structural transition from a low-temperature rhombohedral phase to a high-temperature cubic phase at around 150\,K~\cite{Sayetat:1984}. As in our model the umbrella state is induced by the presence of competing axes, its appearance does not involve any spontaneous symmetry breaking. Therefore, in the framework of this model, the non-collinear umbrella state persists up to the magnetic ordering temperature and some other mechanism would be needed to produce a collinear to non-collinear transition within the ordered phase.
       
\begin{figure}[tb]
\centering
\includegraphics[width=.88\linewidth]{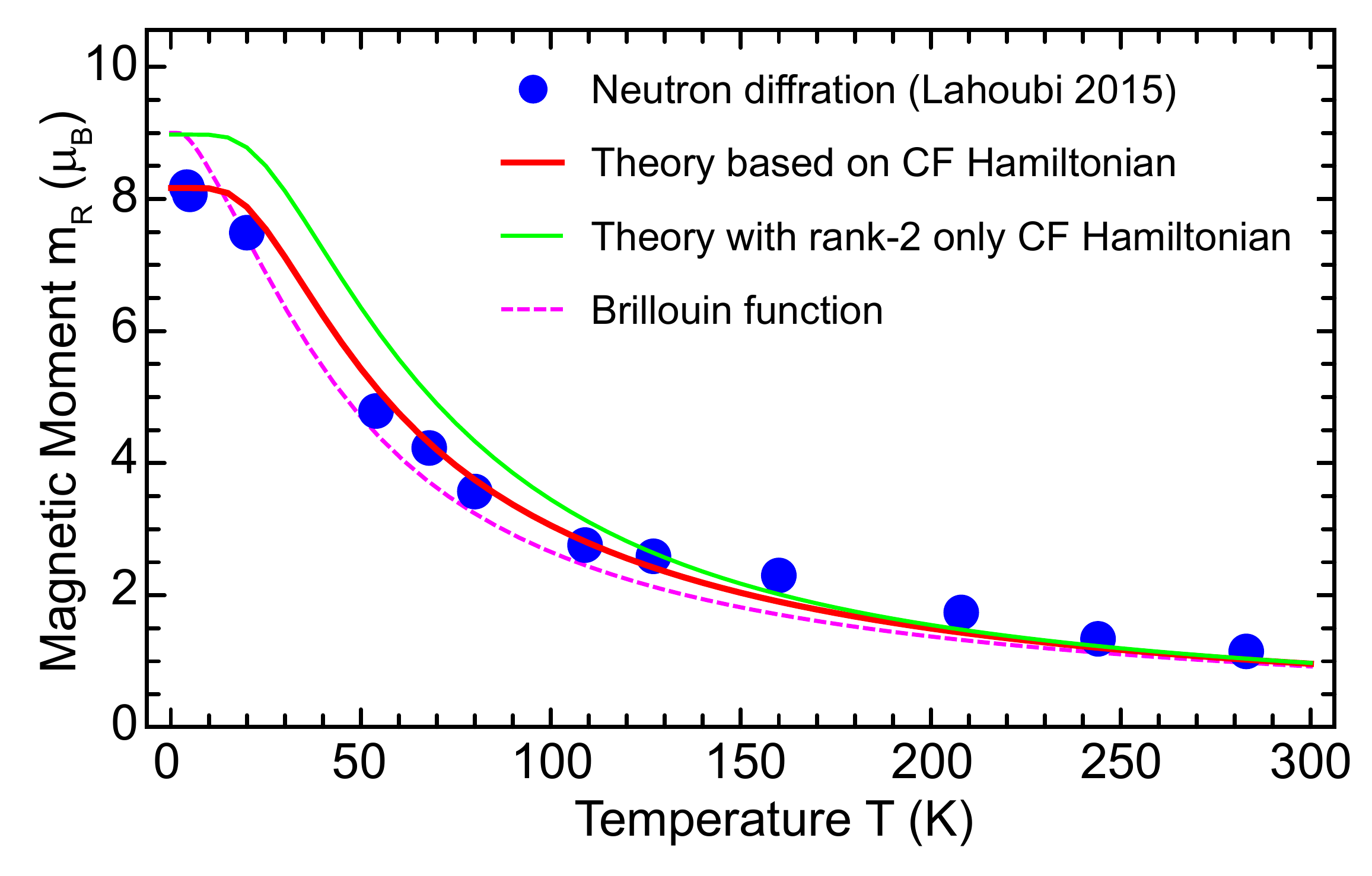} \hspace{0mm}
\caption{Temperature dependence of the magnetic moment of Tb$^{3+}$ in Tb$_{3}$Fe$_{5}$O$_{12}$ (total angular momentum $J=6$ and $g_J = 3/2$). Neutron diffraction measurements (blue dots, Refs.~\cite{Lahoubi:2012, Lahoubi:2015}) are compared with theoretical calculations. The best fit (red thick line) is reproduced by crystal-field (CF) Hamiltonian~\cite{Hau:1986} perturbed by an effective field of 13 Tesla (from the surrounding Fe-ions). The results of calculations using only rank-2 Stevens operators from the same crystal-field Hamiltonian are also shown (green line). The Brillouin function (dashed magenta) uses the mean-field theory in Sec.~\ref{sec:groundstate} with $J_{cd}=0.2J_{ad}$.}
\label{fig:brillouinCF}
\end{figure}

	Using the same parameters, {\it i.e.} $J_{cd}=0.2J_{ad}$ and treating $2 J_{cd} s $ as a temperature independent molecular field from the two next-nearest neighbour Fe-$d$-ions, we calculate a Brillouin function to compare with the temperature dependence of the measured Tb$^{3+}$ magnetic moment in Tb$_{3}$Fe$_{5}$O$_{12}$~\cite{Lahoubi:2012, Lahoubi:2015}. As shown in Figure~\ref{fig:brillouinCF}, we find reasonable agreement, even though at low  temperatures the Brillouin function does not match the measured magnetic moment. This is not surprising as at low temperatures crystal-field effects become more relevant, and the Tb$^{3+}$-ion is prone to a reduction of the magnetic moment~\cite{Wawrzynczak:2019}. 

	To enrich our analysis and include such effects, we then consider the full crystal-field Hamiltonian $\hat{\mathcal{H}}_{\rm{CF}}$ (parameters from Ref.~\cite{Hau:1986}), and we study the behaviour of the magnetic moment of Tb$^{3+}$ at finite temperatures under an effective magnetic perturbation, $\mathbf{B}_{\rm eff}$, mimicking the molecular field on the single rare-earth ion. From exact diagonalisation of $\hat{\mathcal{H}}_{\rm{CF}} - g_{J} \mu_{\rm B} \hat{\mathbf{J}} \cdot \mathbf{B}_{\rm eff}$ we define the micro-canonical ensemble of the single-ion states, and we calculate the thermal average of the magnetic moment using Boltzmann statistics. Tailoring this to our case of interest, the field is taken along the crystallographic $[111]$ direction (the direction of the iron moments neglecting any secondary canting effects), and the thermal expectation value of the single-ion magnetic moment under $\mathbf{B}_{111}$ is also averaged over the six inequivalent coordinate systems~\cite{Wawrzynczak:2019} of the crystallographic $c$-sites of the garnet lattice. As shown by the red line in Fig.~\ref{fig:brillouinCF},  ${B}_{111} =13$\,T accounts for a remarkable overlap with the experimental measurements. Moreover, without any further tuning of parameters, the fit is maintained for the lowest temperatures, in contrast to the Brillouin function.

	In this context,  the existence of actually six different local coordinate systems~\cite{Wawrzynczak:2019} for the rare-earth $c$-sites is not strongly relevant to the curves. This is because the $[111]$ direction is highly symmetric for a cubic lattice and, more importantly, because in Fig.~\ref{fig:brillouinCF} we only average the modulus of the magnetic moment, without considering the local anisotropies. Nevertheless, it should be  emphasized that the crystalline environments cause each local reference frame of a $R^{3+}$-ion in a garnet to ``see differently'' a field pointing along a global crystallographic direction. In fact, even a field aligned with the global $[111]$ direction, would be ``seen'' along the local $\left[1,0,\sqrt{2}\right]$ for the sites $i=1,2,3$, and along the local $\left[-1,0,-\sqrt{2}\right]$ for the sites $i'=4,5,6$ -- see appendix of Ref.~\cite{Wawrzynczak:2019} for the notation of the 6 inequivalent coordinate systems. In general ${\bf x}_{i}=-{\bf x}_{i+3}, {\bf y}_{i}=-{\bf z}_{i+3}, {\bf z}_{i}=-{\bf y}_{i+3}$, with $i=1,2,3$. As discussed in the following section, this distinction is important in that is enough to suggest a simple, and yet original, interpretation of the {\it double} umbrella structure.

\section{Hard-axis anisotropy and the double umbrella structure}
\label{sec:hardaxisdoubleumbrella}

	We now return to the question of the ``double-umbrella" spin structure and its relation to crystal symmetry. In the past it has not been clear whether this complication to the magnetic structure was because of extra features, for example, the lowering of the crystal symmetry from cubic to rhombohedral or coupling to the elastic degrees of freedom. In fact we can now argue that it arises even naturally in the cubic structure, if we include an additional uniaxial magnetic anisotropy with constant $K_h>0$, which results in a magnetically hard axis $\beta_i$. As we know from the crystal-field $D_{2}$ symmetry in garnets, $\beta_i$ lies orthogonal to the easy axis $\alpha_i$ of the $i$-th rare-earth site (see Eq.~\eqref{eq:HDKJ1}). 
	
	The point is that in the cubic garnet structure there are not three but {\it six} distinct sites of the rare-earth ions with 3 {\rm pairs} including the same orthogonal axes, which we have argued leads to the umbrella spin structure. However, as discussed at the end of Sec.~\ref{sec:magneticmomentreal}, the pairs differ in that the local environment is related by a triple of axes which are not simply reversed. For example, one such pair has local axes ${\bf x}_1 = [0,0,1], {\bf y}_1=\frac {1}{\sqrt{2}} [1,\bar{1},0], {\bf{z}}_1=\frac{1}{\sqrt{2}} [1,1,0] $ and $ {\bf x}_4 = [0,0,\bar{1}],  {\bf{ y}}_4=\frac {1}{\sqrt{2}} [\bar{1},\bar{1},0],  {\bf{z}}_4=\frac{1}{\sqrt{2}} [\bar{1},1,0] $, respectively, in the notation of Ref.~\cite{Wawrzynczak:2019}. The other two pairs have coordinates permuted in the three crystal directions. As the single-ion anisotropies are even powers of the spin operators -- they do not break time-reversal symmetry -- the three axes taken as directions without senses are identical, but the order of the second and third axes is reversed: $S_{{\bf y}_4}^2 =S_{{\bf z}_1}^2$, $S_{{\bf z}_4}^2 =S_{{\bf y}_1}^2$. To be consistent with crystal symmetry, the directions of the hard axis, for example, at a site~``1'' will then be orthogonal to the hard axis at site~``4''. In other words, the choice of the [1,1,1] direction for the iron sublattice magnetization breaks symmetry differently with respect to the local axes of the pairs of rare-earth sites: one of the magnetically hard axis $\beta_i$ might be orthogonal to the canting plane and can alter the spin-wave dynamics but not the canting angle as calculated above. On the other site there would be a term of the same numerical coefficient but with axis $\beta^{\prime} \neq \beta$ in the canting plane. The presence of such a term will increase, on half the sites, the effective value of $K_{e}$ in Eq.~\eqref{eq:sublatticeenergy8}, and so it would increase the predicted canting angle on half the triangles. 
	
	We therefore have a simple model of the ``double-umbrella" spin structure by inclusion of a magnetically hard axis. The difference of the angles of the two different umbrellas is thus a measure of the strength of the hard axis. We will not pursue the dynamics of the more complex, double umbrella structure here, as we would have to make our simple model of the lattice rather complex. It is clear that this will lead to six, rather than three, bands that are primarily of single-ion ``crystal-field'' nature with rather different chiral properties.

\section{Conclusions} 
\label{sec:conclusions}

	In conclusion,  we have shown that the simple Hamiltonian ${\cal H}^{0}$ in Eq.~\eqref{eq:HDKJ0}, with or without the extra terms of ${\cal H}^{1}$ in Eq.~\eqref{eq:HDKJ1}, provides a natural description of non-collinear spin structures featuring ``umbrella-like" spin arrangements in rare-earth iron garnets $R_{3}$Fe$_{5}$O$_{12}$. 
	We have demonstrated that this canted spin structure persists up to the magnetic ordering temperature of $R_{3}$Fe$_{5}$O$_{12}$. We further predict an induced canting on the Fe-d moments, which has not been observed experimentally so far. We note that the predicted angles are within past experimental errors quoted~\cite{Sayetat:1984} so this prediction is not excluded. 
	
	We use linear spin-wave theory to calculate the spin-wave dispersion of $R_{3}$Fe$_{5}$O$_{12}$. In the framework of our simple model, the acoustic mode becomes gapped, and its chirality gets mixed. To estimate the quantitative dependence of the gap on the exchange parameters, 
	it is needed an extended three-dimensional model with the iron-iron exchange couplings, $J_{dd}$ and $J_{aa}$, which partially frustrate the ferrimagnetic order.
	The chirality effects are mainly caused by the non-collinearity of the spin structure, which mixes polarisations via hybridization between the propagating parts on the Fe sublattices. This hybridization, produced by the coupling between the rare-earth and the iron moments, has been studied experimentally in Yb$_{3}$Fe$_{5}$O$_{12}$ in light of a microscopic model for the R-Fe exchange interactions~\cite{Pecanha-Antonio:2022}.
	We would expect that inclusion of hard axes will enhance the effects on chirality. The complex chirality of the acoustic mode in Figures \ref{fig:spinwaves1}-\ref{fig:spinwaves2} is clearly different from previous spin-wave calculations of Gd$_{3}$Fe$_{5}$O$_{12}$ with negligible crystal-field effects~\cite{Geprags:2016, Shen:2019} and is expected to have great influence on experiments in the fields of spintronics and spin caloritronics.  For example, both the gapping of the acoustic modes and their depolarisation will most certainly reduce their spin-current contributions to the total spin Seebeck effect signal at low temperatures. However, inelastic neutron scattering experiments with polarisation analysis are needed to attain a full quantitative understanding of the spin-wave dispersion of rare-earth iron garnets at low temperatures.

\section*{Acknowledgments}

	T.~Z. would like to acknowledge the privilege of having met  Igor Dzyaloshinskii in person, thanks to the life-long friendship of Igor with the late Philippe Nozi\`eres. 
	
	T.~Z. acknowledges support for this work from ICC-IMR, Tohoku University. 
	B.~T. would like to thank the theory group of the Institut Laue-Langevin for the hospitality and the support offered for the writing of this manuscript. This publication is funded in part by a QuantEmX grant from ICAM and the Gordon and Betty Moore Foundation through Grant GBMF5305 to Bruno Tomasello.


\begin{thebibliography}{40}%
\makeatletter
\providecommand \@ifxundefined [1]{%
 \@ifx{#1\undefined}
}%
\providecommand \@ifnum [1]{%
 \ifnum #1\expandafter \@firstoftwo
 \else \expandafter \@secondoftwo
 \fi
}%
\providecommand \@ifx [1]{%
 \ifx #1\expandafter \@firstoftwo
 \else \expandafter \@secondoftwo
 \fi
}%
\providecommand \natexlab [1]{#1}%
\providecommand \enquote  [1]{``#1''}%
\providecommand \bibnamefont  [1]{#1}%
\providecommand \bibfnamefont [1]{#1}%
\providecommand \citenamefont [1]{#1}%
\providecommand \href@noop [0]{\@secondoftwo}%
\providecommand \href [0]{\begingroup \@sanitize@url \@href}%
\providecommand \@href[1]{\@@startlink{#1}\@@href}%
\providecommand \@@href[1]{\endgroup#1\@@endlink}%
\providecommand \@sanitize@url [0]{\catcode `\\12\catcode `\$12\catcode
  `\&12\catcode `\#12\catcode `\^12\catcode `\_12\catcode `\%12\relax}%
\providecommand \@@startlink[1]{}%
\providecommand \@@endlink[0]{}%
\providecommand \url  [0]{\begingroup\@sanitize@url \@url }%
\providecommand \@url [1]{\endgroup\@href {#1}{\urlprefix }}%
\providecommand \urlprefix  [0]{URL }%
\providecommand \Eprint [0]{\href }%
\providecommand \doibase [0]{http://dx.doi.org/}%
\providecommand \selectlanguage [0]{\@gobble}%
\providecommand \bibinfo  [0]{\@secondoftwo}%
\providecommand \bibfield  [0]{\@secondoftwo}%
\providecommand \translation [1]{[#1]}%
\providecommand \BibitemOpen [0]{}%
\providecommand \bibitemStop [0]{}%
\providecommand \bibitemNoStop [0]{.\EOS\space}%
\providecommand \EOS [0]{\spacefactor3000\relax}%
\providecommand \BibitemShut  [1]{\csname bibitem#1\endcsname}%
\let\auto@bib@innerbib\@empty
\bibitem [{\citenamefont {Kruglyak}\ \emph {et~al.}(2010)\citenamefont
  {Kruglyak}, \citenamefont {Demokritov},\ and\ \citenamefont
  {Grundler}}]{Kruglyak:2010}%
  \BibitemOpen
  \bibfield  {author} {\bibinfo {author} {\bibfnamefont {V.~V.}\ \bibnamefont
  {Kruglyak}}, \bibinfo {author} {\bibfnamefont {S.~O.}\ \bibnamefont
  {Demokritov}}, \ and\ \bibinfo {author} {\bibfnamefont {D.}~\bibnamefont
  {Grundler}},\ }\href {\doibase 10.1088/0022-3727/43/26/264001} {\bibfield
  {journal} {\bibinfo  {journal} {J. Phys. D: Appl. Phys.}\ }\textbf {\bibinfo
  {volume} {43}},\ \bibinfo {pages} {264001} (\bibinfo {year}
  {2010})}\BibitemShut {NoStop}%
\bibitem [{\citenamefont {Bauer}\ \emph {et~al.}(2012)\citenamefont {Bauer},
  \citenamefont {Saitoh},\ and\ \citenamefont {van Wees}}]{Bauer:2012}%
  \BibitemOpen
  \bibfield  {author} {\bibinfo {author} {\bibfnamefont {G.~E.~W.}\
  \bibnamefont {Bauer}}, \bibinfo {author} {\bibfnamefont {E.}~\bibnamefont
  {Saitoh}}, \ and\ \bibinfo {author} {\bibfnamefont {B.~J.}\ \bibnamefont {van
  Wees}},\ }\href {\doibase 10.1038/nmat3301} {\bibfield  {journal} {\bibinfo
  {journal} {Nat. Mater.}\ }\textbf {\bibinfo {volume} {11}},\ \bibinfo {pages}
  {391} (\bibinfo {year} {2012})}\BibitemShut {NoStop}%
\bibitem [{\citenamefont {Han}\ \emph {et~al.}(2018)\citenamefont {Han},
  \citenamefont {Otani},\ and\ \citenamefont {Maekawa}}]{Han:2018}%
  \BibitemOpen
  \bibfield  {author} {\bibinfo {author} {\bibfnamefont {W.}~\bibnamefont
  {Han}}, \bibinfo {author} {\bibfnamefont {Y.}~\bibnamefont {Otani}}, \ and\
  \bibinfo {author} {\bibfnamefont {S.}~\bibnamefont {Maekawa}},\ }\href@noop
  {} {\bibfield  {journal} {\bibinfo  {journal} {npj Quantum Materials}\
  }\textbf {\bibinfo {volume} {3}},\ \bibinfo {pages} {27} (\bibinfo {year}
  {2018})}\BibitemShut {NoStop}%
\bibitem [{\citenamefont {Brataas}\ \emph {et~al.}(2020)\citenamefont
  {Brataas}, \citenamefont {{van Wees}}, \citenamefont {Klein}, \citenamefont
  {{de Loubens}},\ and\ \citenamefont {Viret}}]{Brataas:2020}%
  \BibitemOpen
  \bibfield  {author} {\bibinfo {author} {\bibfnamefont {A.}~\bibnamefont
  {Brataas}}, \bibinfo {author} {\bibfnamefont {B.}~\bibnamefont {{van Wees}}},
  \bibinfo {author} {\bibfnamefont {O.}~\bibnamefont {Klein}}, \bibinfo
  {author} {\bibfnamefont {G.}~\bibnamefont {{de Loubens}}}, \ and\ \bibinfo
  {author} {\bibfnamefont {M.}~\bibnamefont {Viret}},\ }\href {\doibase
  https://doi.org/10.1016/j.physrep.2020.08.006} {\bibfield  {journal}
  {\bibinfo  {journal} {Phys. Rep.}\ }\textbf {\bibinfo {volume} {885}},\
  \bibinfo {pages} {1} (\bibinfo {year} {2020})},\ \bibinfo {note} {spin
  Insulatronics}\BibitemShut {NoStop}%
\bibitem [{\citenamefont {Cornelissen}\ \emph {et~al.}(2015)\citenamefont
  {Cornelissen}, \citenamefont {Liu}, \citenamefont {Duine}, \citenamefont
  {Youssef},\ and\ \citenamefont {van Wees}}]{Cornelissen:2015}%
  \BibitemOpen
  \bibfield  {author} {\bibinfo {author} {\bibfnamefont {L.~J.}\ \bibnamefont
  {Cornelissen}}, \bibinfo {author} {\bibfnamefont {J.}~\bibnamefont {Liu}},
  \bibinfo {author} {\bibfnamefont {R.~A.}\ \bibnamefont {Duine}}, \bibinfo
  {author} {\bibfnamefont {J.~B.}\ \bibnamefont {Youssef}}, \ and\ \bibinfo
  {author} {\bibfnamefont {B.~J.}\ \bibnamefont {van Wees}},\ }\href
  {https://doi.org/10.1038/nphys3465} {\bibfield  {journal} {\bibinfo
  {journal} {Nature Physics}\ }\textbf {\bibinfo {volume} {11}},\ \bibinfo
  {pages} {1022} (\bibinfo {year} {2015})}\BibitemShut {NoStop}%
\bibitem [{\citenamefont {Jamison}\ \emph {et~al.}(2019)\citenamefont
  {Jamison}, \citenamefont {Yang}, \citenamefont {Giles}, \citenamefont
  {Brangham}, \citenamefont {Wu}, \citenamefont {Hammel}, \citenamefont
  {Yang},\ and\ \citenamefont {Myers}}]{Jamison:2019}%
  \BibitemOpen
  \bibfield  {author} {\bibinfo {author} {\bibfnamefont {J.~S.}\ \bibnamefont
  {Jamison}}, \bibinfo {author} {\bibfnamefont {Z.}~\bibnamefont {Yang}},
  \bibinfo {author} {\bibfnamefont {B.~L.}\ \bibnamefont {Giles}}, \bibinfo
  {author} {\bibfnamefont {J.~T.}\ \bibnamefont {Brangham}}, \bibinfo {author}
  {\bibfnamefont {G.}~\bibnamefont {Wu}}, \bibinfo {author} {\bibfnamefont
  {P.~C.}\ \bibnamefont {Hammel}}, \bibinfo {author} {\bibfnamefont
  {F.}~\bibnamefont {Yang}}, \ and\ \bibinfo {author} {\bibfnamefont {R.~C.}\
  \bibnamefont {Myers}},\ }\href {\doibase 10.1103/PhysRevB.100.134402}
  {\bibfield  {journal} {\bibinfo  {journal} {Phys. Rev. B}\ }\textbf {\bibinfo
  {volume} {100}},\ \bibinfo {pages} {134402} (\bibinfo {year}
  {2019})}\BibitemShut {NoStop}%
\bibitem [{\citenamefont {Serga}\ \emph {et~al.}(2010)\citenamefont {Serga},
  \citenamefont {Chumak},\ and\ \citenamefont {Hillebrands}}]{Serga:2010}%
  \BibitemOpen
  \bibfield  {author} {\bibinfo {author} {\bibfnamefont {A.~A.}\ \bibnamefont
  {Serga}}, \bibinfo {author} {\bibfnamefont {A.~V.}\ \bibnamefont {Chumak}}, \
  and\ \bibinfo {author} {\bibfnamefont {B.}~\bibnamefont {Hillebrands}},\
  }\href {\doibase 10.1088/0022-3727/43/26/264002} {\bibfield  {journal}
  {\bibinfo  {journal} {J. Phys. D: Appl. Phys.}\ }\textbf {\bibinfo {volume}
  {43}},\ \bibinfo {pages} {264002} (\bibinfo {year} {2010})}\BibitemShut
  {NoStop}%
\bibitem [{\citenamefont {Newman}\ and\ \citenamefont
  {Stedman}(1969)}]{Newman:1969}%
  \BibitemOpen
  \bibfield  {author} {\bibinfo {author} {\bibfnamefont {D.~J.}\ \bibnamefont
  {Newman}}\ and\ \bibinfo {author} {\bibfnamefont {G.~E.}\ \bibnamefont
  {Stedman}},\ }\href {\doibase 10.1063/1.1672450} {\bibfield  {journal}
  {\bibinfo  {journal} {The Journal of Chemical Physics}\ }\textbf {\bibinfo
  {volume} {51}},\ \bibinfo {pages} {3013} (\bibinfo {year}
  {1969})}\BibitemShut {NoStop}%
\bibitem [{\citenamefont {N{\'e}el}\ \emph {et~al.}(1964)\citenamefont
  {N{\'e}el}, \citenamefont {Pauthenet},\ and\ \citenamefont
  {Dreyfus}}]{Neel:1964}%
  \BibitemOpen
  \bibfield  {author} {\bibinfo {author} {\bibfnamefont {L.}~\bibnamefont
  {N{\'e}el}}, \bibinfo {author} {\bibfnamefont {R.}~\bibnamefont {Pauthenet}},
  \ and\ \bibinfo {author} {\bibfnamefont {B.}~\bibnamefont {Dreyfus}},\
  }\enquote {\bibinfo {title} {Chapter vii the rare earth garnets},}\ in\ \href
  {\doibase https://doi.org/10.1016/S0079-6417(08)60155-9} {\emph {\bibinfo
  {booktitle} {Progress in Low Temperature Physics}}},\ Vol.~\bibinfo {volume}
  {4},\ \bibinfo {editor} {edited by\ \bibinfo {editor} {\bibfnamefont {C.~J.}\
  \bibnamefont {Gorter}}}\ (\bibinfo  {publisher} {Elsevier},\ \bibinfo {year}
  {1964})\ pp.\ \bibinfo {pages} {344--383}\BibitemShut {NoStop}%
\bibitem [{\citenamefont {Pauthenet}(1958)}]{Pauthenet:1958}%
  \BibitemOpen
  \bibfield  {author} {\bibinfo {author} {\bibfnamefont {R.}~\bibnamefont
  {Pauthenet}},\ }\href {\doibase 10.1063/1.1723094} {\bibfield  {journal}
  {\bibinfo  {journal} {J. Appl. Phys.}\ }\textbf {\bibinfo {volume} {29}},\
  \bibinfo {pages} {253} (\bibinfo {year} {1958})}\BibitemShut {NoStop}%
\bibitem [{\citenamefont {N{\'e}el}(1954)}]{Neel:1954}%
  \BibitemOpen
  \bibfield  {author} {\bibinfo {author} {\bibfnamefont {L.}~\bibnamefont
  {N{\'e}el}},\
  }\href@noop {} {\bibfield  {journal} {\bibinfo  {journal} {Comptes Rendus de
  l'Acad{\'e}mie des Sciences}\ }\textbf {\bibinfo {volume} {239}} (\bibinfo
  {year} {1954})}\BibitemShut {NoStop}%
\bibitem [{\citenamefont {Belov}(1996)}]{Belov:1996}%
  \BibitemOpen
  \bibfield  {author} {\bibinfo {author} {\bibfnamefont {K.~P.}\ \bibnamefont
  {Belov}},\ }\href {http://stacks.iop.org/1063-7869/39/i=6/a=R06} {\bibfield
  {journal} {\bibinfo  {journal} {Phys. Usp.}\ }\textbf {\bibinfo {volume}
  {39}},\ \bibinfo {pages} {623} (\bibinfo {year} {1996})}\BibitemShut
  {NoStop}%
\bibitem [{\citenamefont {Geller}\ and\ \citenamefont
  {Gilleo}(1957)}]{Geller:1957a}%
  \BibitemOpen
  \bibfield  {author} {\bibinfo {author} {\bibfnamefont {S.}~\bibnamefont
  {Geller}}\ and\ \bibinfo {author} {\bibfnamefont {M.}~\bibnamefont
  {Gilleo}},\ }\href {\doibase https://doi.org/10.1016/0022-3697(57)90044-6}
  {\bibfield  {journal} {\bibinfo  {journal} {Journal of Physics and Chemistry
  of Solids}\ }\textbf {\bibinfo {volume} {3}},\ \bibinfo {pages} {30}
  (\bibinfo {year} {1957})}\BibitemShut {NoStop}%
\bibitem [{\citenamefont {Tcheou}\ \emph {et~al.}(1970)\citenamefont {Tcheou},
  \citenamefont {Bertaut},\ and\ \citenamefont {Fuess}}]{Tcheou:1970}%
  \BibitemOpen
  \bibfield  {author} {\bibinfo {author} {\bibfnamefont {F.}~\bibnamefont
  {Tcheou}}, \bibinfo {author} {\bibfnamefont {E.~F.}\ \bibnamefont {Bertaut}},
  \ and\ \bibinfo {author} {\bibfnamefont {H.}~\bibnamefont {Fuess}},\ }\href
  {\doibase https://doi.org/10.1016/0038-1098(70)90389-3} {\bibfield  {journal}
  {\bibinfo  {journal} {Solid State Communications}\ }\textbf {\bibinfo
  {volume} {8}},\ \bibinfo {pages} {1751} (\bibinfo {year} {1970})}\BibitemShut
  {NoStop}%
\bibitem [{\citenamefont {Pickart}\ \emph {et~al.}(1970)\citenamefont
  {Pickart}, \citenamefont {Alperin},\ and\ \citenamefont
  {Clark}}]{Pickart:1970}%
  \BibitemOpen
  \bibfield  {author} {\bibinfo {author} {\bibfnamefont {S.~J.}\ \bibnamefont
  {Pickart}}, \bibinfo {author} {\bibfnamefont {H.~A.}\ \bibnamefont
  {Alperin}}, \ and\ \bibinfo {author} {\bibfnamefont {A.~E.}\ \bibnamefont
  {Clark}},\ }\href {\doibase 10.1063/1.1658874} {\bibfield  {journal}
  {\bibinfo  {journal} {J. Appl. Phys.}\ }\textbf {\bibinfo {volume} {41}},\
  \bibinfo {pages} {1192} (\bibinfo {year} {1970})}\BibitemShut {NoStop}%
\bibitem [{\citenamefont {Guillot}\ \emph {et~al.}(1982)\citenamefont
  {Guillot}, \citenamefont {Marchand}, \citenamefont {Tch{\'e}ou},\ and\
  \citenamefont {Feldmann}}]{Guillot:1982}%
  \BibitemOpen
  \bibfield  {author} {\bibinfo {author} {\bibfnamefont {M.}~\bibnamefont
  {Guillot}}, \bibinfo {author} {\bibfnamefont {A.}~\bibnamefont {Marchand}},
  \bibinfo {author} {\bibfnamefont {F.}~\bibnamefont {Tch{\'e}ou}}, \ and\
  \bibinfo {author} {\bibfnamefont {P.}~\bibnamefont {Feldmann}},\ }\href
  {\doibase 10.1063/1.330943} {\bibfield  {journal} {\bibinfo  {journal}
  {Journal of Applied Physics}\ }\textbf {\bibinfo {volume} {53}},\ \bibinfo
  {pages} {2719} (\bibinfo {year} {1982})}\BibitemShut {NoStop}%
\bibitem [{\citenamefont {Lahoubi}\ \emph {et~al.}(1984)\citenamefont
  {Lahoubi}, \citenamefont {Guillot}, \citenamefont {Marchand}, \citenamefont
  {Tcheou},\ and\ \citenamefont {Roudault}}]{Lahoubi:1984}%
  \BibitemOpen
  \bibfield  {author} {\bibinfo {author} {\bibfnamefont {M.}~\bibnamefont
  {Lahoubi}}, \bibinfo {author} {\bibfnamefont {M.}~\bibnamefont {Guillot}},
  \bibinfo {author} {\bibfnamefont {A.}~\bibnamefont {Marchand}}, \bibinfo
  {author} {\bibfnamefont {F.}~\bibnamefont {Tcheou}}, \ and\ \bibinfo {author}
  {\bibfnamefont {E.}~\bibnamefont {Roudault}},\ }\href {\doibase
  10.1109/TMAG.1984.1063259} {\bibfield  {journal} {\bibinfo  {journal} {IEEE
  Trans. Magn.}\ }\textbf {\bibinfo {volume} {20}},\ \bibinfo {pages} {1518}
  (\bibinfo {year} {1984})}\BibitemShut {NoStop}%
\bibitem [{\citenamefont {Hock}\ \emph {et~al.}(1990)\citenamefont {Hock},
  \citenamefont {Fuess}, \citenamefont {Vogt},\ and\ \citenamefont
  {Bonnet}}]{Hock:1990}%
  \BibitemOpen
  \bibfield  {author} {\bibinfo {author} {\bibfnamefont {R.}~\bibnamefont
  {Hock}}, \bibinfo {author} {\bibfnamefont {H.}~\bibnamefont {Fuess}},
  \bibinfo {author} {\bibfnamefont {T.}~\bibnamefont {Vogt}}, \ and\ \bibinfo
  {author} {\bibfnamefont {M.}~\bibnamefont {Bonnet}},\ }\href {\doibase
  https://doi.org/10.1016/0022-4596(90)90182-W} {\bibfield  {journal} {\bibinfo
   {journal} {J. Solid State Chem.}\ }\textbf {\bibinfo {volume} {84}},\
  \bibinfo {pages} {39} (\bibinfo {year} {1990})}\BibitemShut {NoStop}%
\bibitem [{\citenamefont {Princep}\ \emph {et~al.}(2013)\citenamefont
  {Princep}, \citenamefont {Prabhakaran}, \citenamefont {Boothroyd},\ and\
  \citenamefont {Adroja}}]{Princep:2013}%
  \BibitemOpen
  \bibfield  {author} {\bibinfo {author} {\bibfnamefont {A.~J.}\ \bibnamefont
  {Princep}}, \bibinfo {author} {\bibfnamefont {D.}~\bibnamefont
  {Prabhakaran}}, \bibinfo {author} {\bibfnamefont {A.~T.}\ \bibnamefont
  {Boothroyd}}, \ and\ \bibinfo {author} {\bibfnamefont {D.~T.}\ \bibnamefont
  {Adroja}},\ }\href {http://link.aps.org/doi/10.1103/PhysRevB.88.104421}
  {\bibfield  {journal} {\bibinfo  {journal} {Physical Review B}\ }\textbf
  {\bibinfo {volume} {88}},\ \bibinfo {pages} {104421} (\bibinfo {year}
  {2013})}\BibitemShut {NoStop}%
\bibitem [{\citenamefont {Barker}\ and\ \citenamefont
  {Bauer}(2016)}]{Barker:2016}%
  \BibitemOpen
  \bibfield  {author} {\bibinfo {author} {\bibfnamefont {J.}~\bibnamefont
  {Barker}}\ and\ \bibinfo {author} {\bibfnamefont {G.~E.~W.}\ \bibnamefont
  {Bauer}},\ }\href {\doibase 10.1103/PhysRevLett.117.217201} {\bibfield
  {journal} {\bibinfo  {journal} {Physical Review Letters}\ }\textbf {\bibinfo
  {volume} {117}},\ \bibinfo {pages} {217201} (\bibinfo {year}
  {2016})}\BibitemShut {NoStop}%
\bibitem [{\citenamefont {Shen}(2019)}]{Shen:2019}%
  \BibitemOpen
  \bibfield  {author} {\bibinfo {author} {\bibfnamefont {K.}~\bibnamefont
  {Shen}},\ }\href {\doibase 10.1103/PhysRevB.99.024417} {\bibfield  {journal}
  {\bibinfo  {journal} {Physical Review B}\ }\textbf {\bibinfo {volume} {99}},\
  \bibinfo {pages} {024417} (\bibinfo {year} {2019})}\BibitemShut {NoStop}%
\bibitem [{\citenamefont {Shamoto}\ \emph {et~al.}(2020)\citenamefont
  {Shamoto}, \citenamefont {Yasui}, \citenamefont {Matsuura}, \citenamefont
  {Akatsu}, \citenamefont {Kobayashi}, \citenamefont {Nemoto},\ and\
  \citenamefont {Ieda}}]{Shamoto:2020}%
  \BibitemOpen
  \bibfield  {author} {\bibinfo {author} {\bibfnamefont {S.-i.}\ \bibnamefont
  {Shamoto}}, \bibinfo {author} {\bibfnamefont {Y.}~\bibnamefont {Yasui}},
  \bibinfo {author} {\bibfnamefont {M.}~\bibnamefont {Matsuura}}, \bibinfo
  {author} {\bibfnamefont {M.}~\bibnamefont {Akatsu}}, \bibinfo {author}
  {\bibfnamefont {Y.}~\bibnamefont {Kobayashi}}, \bibinfo {author}
  {\bibfnamefont {Y.}~\bibnamefont {Nemoto}}, \ and\ \bibinfo {author}
  {\bibfnamefont {J.}~\bibnamefont {Ieda}},\ }\href {\doibase
  10.1103/PhysRevResearch.2.033235} {\bibfield  {journal} {\bibinfo  {journal}
  {Phys. Rev. Research}\ }\textbf {\bibinfo {volume} {2}},\ \bibinfo {pages}
  {033235} (\bibinfo {year} {2020})}\BibitemShut {NoStop}%
\bibitem [{\citenamefont {Nambu}\ \emph {et~al.}(2020)\citenamefont {Nambu},
  \citenamefont {Barker}, \citenamefont {Okino}, \citenamefont {Kikkawa},
  \citenamefont {Shiomi}, \citenamefont {Enderle}, \citenamefont {Weber},
  \citenamefont {Winn}, \citenamefont {Graves-Brook}, \citenamefont
  {Tranquada}, \citenamefont {Ziman}, \citenamefont {Fujita}, \citenamefont
  {Bauer}, \citenamefont {Saitoh},\ and\ \citenamefont {Kakurai}}]{Nambu:2020}%
  \BibitemOpen
  \bibfield  {author} {\bibinfo {author} {\bibfnamefont {Y.}~\bibnamefont
  {Nambu}}, \bibinfo {author} {\bibfnamefont {J.}~\bibnamefont {Barker}},
  \bibinfo {author} {\bibfnamefont {Y.}~\bibnamefont {Okino}}, \bibinfo
  {author} {\bibfnamefont {T.}~\bibnamefont {Kikkawa}}, \bibinfo {author}
  {\bibfnamefont {Y.}~\bibnamefont {Shiomi}}, \bibinfo {author} {\bibfnamefont
  {M.}~\bibnamefont {Enderle}}, \bibinfo {author} {\bibfnamefont
  {T.}~\bibnamefont {Weber}}, \bibinfo {author} {\bibfnamefont
  {B.}~\bibnamefont {Winn}}, \bibinfo {author} {\bibfnamefont {M.}~\bibnamefont
  {Graves-Brook}}, \bibinfo {author} {\bibfnamefont {J.~M.}\ \bibnamefont
  {Tranquada}}, \bibinfo {author} {\bibfnamefont {T.}~\bibnamefont {Ziman}},
  \bibinfo {author} {\bibfnamefont {M.}~\bibnamefont {Fujita}}, \bibinfo
  {author} {\bibfnamefont {G.~E.~W.}\ \bibnamefont {Bauer}}, \bibinfo {author}
  {\bibfnamefont {E.}~\bibnamefont {Saitoh}}, \ and\ \bibinfo {author}
  {\bibfnamefont {K.}~\bibnamefont {Kakurai}},\ }\href {\doibase
  10.1103/PhysRevLett.125.027201} {\bibfield  {journal} {\bibinfo  {journal}
  {Phys. Rev. Lett.}\ }\textbf {\bibinfo {volume} {125}},\ \bibinfo {pages}
  {027201} (\bibinfo {year} {2020})}\BibitemShut {NoStop}%
\bibitem [{\citenamefont {Gepr{\"a}gs}\ \emph {et~al.}(2016)\citenamefont
  {Gepr{\"a}gs}, \citenamefont {Kehlberger}, \citenamefont {Coletta},
  \citenamefont {Qiu}, \citenamefont {Guo}, \citenamefont {Schulz},
  \citenamefont {Mix}, \citenamefont {Meyer}, \citenamefont {Kamra},
  \citenamefont {Althammer}, \citenamefont {Huebl}, \citenamefont {Jakob},
  \citenamefont {Ohnuma}, \citenamefont {Adachi}, \citenamefont {Barker},
  \citenamefont {Maekawa}, \citenamefont {Bauer}, \citenamefont {Saitoh},
  \citenamefont {Gross}, \citenamefont {Goennenwein},\ and\ \citenamefont
  {Kl{\"a}ui}}]{Geprags:2016}%
  \BibitemOpen
  \bibfield  {author} {\bibinfo {author} {\bibfnamefont {S.}~\bibnamefont
  {Gepr{\"a}gs}}, \bibinfo {author} {\bibfnamefont {A.}~\bibnamefont
  {Kehlberger}}, \bibinfo {author} {\bibfnamefont {F.~D.}\ \bibnamefont
  {Coletta}}, \bibinfo {author} {\bibfnamefont {Z.}~\bibnamefont {Qiu}},
  \bibinfo {author} {\bibfnamefont {E.-J.}\ \bibnamefont {Guo}}, \bibinfo
  {author} {\bibfnamefont {T.}~\bibnamefont {Schulz}}, \bibinfo {author}
  {\bibfnamefont {C.}~\bibnamefont {Mix}}, \bibinfo {author} {\bibfnamefont
  {S.}~\bibnamefont {Meyer}}, \bibinfo {author} {\bibfnamefont
  {A.}~\bibnamefont {Kamra}}, \bibinfo {author} {\bibfnamefont
  {M.}~\bibnamefont {Althammer}}, \bibinfo {author} {\bibfnamefont
  {H.}~\bibnamefont {Huebl}}, \bibinfo {author} {\bibfnamefont
  {G.}~\bibnamefont {Jakob}}, \bibinfo {author} {\bibfnamefont
  {Y.}~\bibnamefont {Ohnuma}}, \bibinfo {author} {\bibfnamefont
  {H.}~\bibnamefont {Adachi}}, \bibinfo {author} {\bibfnamefont
  {J.}~\bibnamefont {Barker}}, \bibinfo {author} {\bibfnamefont
  {S.}~\bibnamefont {Maekawa}}, \bibinfo {author} {\bibfnamefont {G.~E.~W.}\
  \bibnamefont {Bauer}}, \bibinfo {author} {\bibfnamefont {E.}~\bibnamefont
  {Saitoh}}, \bibinfo {author} {\bibfnamefont {R.}~\bibnamefont {Gross}},
  \bibinfo {author} {\bibfnamefont {S.~T.~B.}\ \bibnamefont {Goennenwein}}, \
  and\ \bibinfo {author} {\bibfnamefont {M.}~\bibnamefont {Kl{\"a}ui}},\ }\href
  {\doibase 10.1038/ncomms10452} {\bibfield  {journal} {\bibinfo  {journal}
  {Nature Communications}\ }\textbf {\bibinfo {volume} {7}},\ \bibinfo {pages}
  {10452} (\bibinfo {year} {2016})}\BibitemShut {NoStop}%
\bibitem [{\citenamefont {Hock}\ \emph {et~al.}(1991)\citenamefont {Hock},
  \citenamefont {Fuess}, \citenamefont {Vogt},\ and\ \citenamefont
  {Bonnet}}]{Hock:1991}%
  \BibitemOpen
  \bibfield  {author} {\bibinfo {author} {\bibfnamefont {R.}~\bibnamefont
  {Hock}}, \bibinfo {author} {\bibfnamefont {H.}~\bibnamefont {Fuess}},
  \bibinfo {author} {\bibfnamefont {T.}~\bibnamefont {Vogt}}, \ and\ \bibinfo
  {author} {\bibfnamefont {M.}~\bibnamefont {Bonnet}},\ }\href {\doibase
  10.1007/BF01324338} {\bibfield  {journal} {\bibinfo  {journal} {Zeitschrift
  f{\"u}r Physik B Condensed Matter}\ }\textbf {\bibinfo {volume} {82}},\
  \bibinfo {pages} {283} (\bibinfo {year} {1991})}\BibitemShut {NoStop}%
\bibitem [{\citenamefont {Lahoubi}(2012{\natexlab{a}})}]{Lahoubi:2012}%
  \BibitemOpen
  \bibfield  {author} {\bibinfo {author} {\bibfnamefont {M.}~\bibnamefont
  {Lahoubi}},\ }\href {\doibase 10.1088/1742-6596/340/1/012068} {\bibfield
  {journal} {\bibinfo  {journal} {Journal of Physics: Conference Series}\
  }\textbf {\bibinfo {volume} {340}},\ \bibinfo {pages} {012068} (\bibinfo
  {year} {2012}{\natexlab{a}})}\BibitemShut {NoStop}%
\bibitem [{\citenamefont {Hau}\ \emph {et~al.}(1986)\citenamefont {Hau},
  \citenamefont {Porcher}, \citenamefont {Vien},\ and\ \citenamefont
  {Pajot}}]{Hau:1986}%
  \BibitemOpen
  \bibfield  {author} {\bibinfo {author} {\bibfnamefont {N.~H.}\ \bibnamefont
  {Hau}}, \bibinfo {author} {\bibfnamefont {P.}~\bibnamefont {Porcher}},
  \bibinfo {author} {\bibfnamefont {T.~K.}\ \bibnamefont {Vien}}, \ and\
  \bibinfo {author} {\bibfnamefont {B.}~\bibnamefont {Pajot}},\ }\href
  {\doibase https://doi.org/10.1016/0022-3697(86)90181-2} {\bibfield  {journal}
  {\bibinfo  {journal} {Journal of Physics and Chemistry of Solids}\ }\textbf
  {\bibinfo {volume} {47}},\ \bibinfo {pages} {83} (\bibinfo {year}
  {1986})}\BibitemShut {NoStop}%
\bibitem [{\citenamefont {Wawrzy{\'n}czak}\ \emph {et~al.}(2019)\citenamefont
  {Wawrzy{\'n}czak}, \citenamefont {Tomasello}, \citenamefont {Manuel},
  \citenamefont {Khalyavin}, \citenamefont {Le}, \citenamefont {Guidi},
  \citenamefont {Cervellino}, \citenamefont {Ziman}, \citenamefont {Boehm},
  \citenamefont {Nilsen},\ and\ \citenamefont {Fennell}}]{Wawrzynczak:2019}%
  \BibitemOpen
  \bibfield  {author} {\bibinfo {author} {\bibfnamefont {R.}~\bibnamefont
  {Wawrzy{\'n}czak}}, \bibinfo {author} {\bibfnamefont {B.}~\bibnamefont
  {Tomasello}}, \bibinfo {author} {\bibfnamefont {P.}~\bibnamefont {Manuel}},
  \bibinfo {author} {\bibfnamefont {D.}~\bibnamefont {Khalyavin}}, \bibinfo
  {author} {\bibfnamefont {M.~D.}\ \bibnamefont {Le}}, \bibinfo {author}
  {\bibfnamefont {T.}~\bibnamefont {Guidi}}, \bibinfo {author} {\bibfnamefont
  {A.}~\bibnamefont {Cervellino}}, \bibinfo {author} {\bibfnamefont
  {T.}~\bibnamefont {Ziman}}, \bibinfo {author} {\bibfnamefont
  {M.}~\bibnamefont {Boehm}}, \bibinfo {author} {\bibfnamefont {G.~J.}\
  \bibnamefont {Nilsen}}, \ and\ \bibinfo {author} {\bibfnamefont
  {T.}~\bibnamefont {Fennell}},\ }\href {\doibase 10.1103/PhysRevB.100.094442}
  {\bibfield  {journal} {\bibinfo  {journal} {Physical Review B}\ }\textbf
  {\bibinfo {volume} {100}},\ \bibinfo {pages} {094442} (\bibinfo {year}
  {2019})}\BibitemShut {NoStop}%
\bibitem [{\citenamefont {Tinkham}(1961)}]{Tinkham:1961}%
  \BibitemOpen
  \bibfield  {author} {\bibinfo {author} {\bibfnamefont {M.}~\bibnamefont
  {Tinkham}},\ }\href {\doibase 10.1103/PhysRev.124.311} {\bibfield  {journal}
  {\bibinfo  {journal} {Physical Review}\ }\textbf {\bibinfo {volume} {124}},\
  \bibinfo {pages} {311} (\bibinfo {year} {1961})}\BibitemShut {NoStop}%
\bibitem [{\citenamefont {Princep}\ \emph {et~al.}(2017)\citenamefont
  {Princep}, \citenamefont {Ewings}, \citenamefont {Ward}, \citenamefont
  {T{\'o}th}, \citenamefont {Dubs}, \citenamefont {Prabhakaran},\ and\
  \citenamefont {Boothroyd}}]{Princep:2017}%
  \BibitemOpen
  \bibfield  {author} {\bibinfo {author} {\bibfnamefont {A.~J.}\ \bibnamefont
  {Princep}}, \bibinfo {author} {\bibfnamefont {R.~A.}\ \bibnamefont {Ewings}},
  \bibinfo {author} {\bibfnamefont {S.}~\bibnamefont {Ward}}, \bibinfo {author}
  {\bibfnamefont {S.}~\bibnamefont {T{\'o}th}}, \bibinfo {author}
  {\bibfnamefont {C.}~\bibnamefont {Dubs}}, \bibinfo {author} {\bibfnamefont
  {D.}~\bibnamefont {Prabhakaran}}, \ and\ \bibinfo {author} {\bibfnamefont
  {A.~T.}\ \bibnamefont {Boothroyd}},\ }\href
  {https://doi.org/10.1038/s41535-017-0067-y} {\bibfield  {journal} {\bibinfo
  {journal} {npj Quantum Materials}\ }\textbf {\bibinfo {volume} {2}},\
  \bibinfo {pages} {63} (\bibinfo {year} {2017})}\BibitemShut {NoStop}%
\bibitem [{\citenamefont {Cherepanov}\ \emph {et~al.}(1993)\citenamefont
  {Cherepanov}, \citenamefont {Kolokolov},\ and\ \citenamefont
  {L'vov}}]{Cherepanov:1993}%
  \BibitemOpen
  \bibfield  {author} {\bibinfo {author} {\bibfnamefont {V.}~\bibnamefont
  {Cherepanov}}, \bibinfo {author} {\bibfnamefont {I.}~\bibnamefont
  {Kolokolov}}, \ and\ \bibinfo {author} {\bibfnamefont {V.}~\bibnamefont
  {L'vov}},\ }\href {\doibase https://doi.org/10.1016/0370-1573(93)90107-O}
  {\bibfield  {journal} {\bibinfo  {journal} {Physics Reports}\ }\textbf
  {\bibinfo {volume} {229}},\ \bibinfo {pages} {81} (\bibinfo {year}
  {1993})}\BibitemShut {NoStop}%
\bibitem [{\citenamefont {Cooper}\ \emph {et~al.}(1962)\citenamefont {Cooper},
  \citenamefont {Elliott}, \citenamefont {Nettel},\ and\ \citenamefont
  {Suhl}}]{Cooper:1962}%
  \BibitemOpen
  \bibfield  {author} {\bibinfo {author} {\bibfnamefont {B.~R.}\ \bibnamefont
  {Cooper}}, \bibinfo {author} {\bibfnamefont {R.~J.}\ \bibnamefont {Elliott}},
  \bibinfo {author} {\bibfnamefont {S.~J.}\ \bibnamefont {Nettel}}, \ and\
  \bibinfo {author} {\bibfnamefont {H.}~\bibnamefont {Suhl}},\ }\href {\doibase
  10.1103/PhysRev.127.57} {\bibfield  {journal} {\bibinfo  {journal} {Phys.
  Rev.}\ }\textbf {\bibinfo {volume} {127}},\ \bibinfo {pages} {57} (\bibinfo
  {year} {1962})}\BibitemShut {NoStop}%
\bibitem [{\citenamefont {Harris}(1963)}]{Harris:1963}%
  \BibitemOpen
  \bibfield  {author} {\bibinfo {author} {\bibfnamefont {A.~B.}\ \bibnamefont
  {Harris}},\ }\href {\doibase 10.1103/PhysRev.132.2398} {\bibfield  {journal}
  {\bibinfo  {journal} {Physical Review}\ }\textbf {\bibinfo {volume} {132}},\
  \bibinfo {pages} {2398} (\bibinfo {year} {1963})}\BibitemShut {NoStop}%
\bibitem [{\citenamefont {Ohnuma}\ \emph {et~al.}(2013)\citenamefont {Ohnuma},
  \citenamefont {Adachi}, \citenamefont {Saitoh},\ and\ \citenamefont
  {Maekawa}}]{Ohnuma:2013}%
  \BibitemOpen
  \bibfield  {author} {\bibinfo {author} {\bibfnamefont {Y.}~\bibnamefont
  {Ohnuma}}, \bibinfo {author} {\bibfnamefont {H.}~\bibnamefont {Adachi}},
  \bibinfo {author} {\bibfnamefont {E.}~\bibnamefont {Saitoh}}, \ and\ \bibinfo
  {author} {\bibfnamefont {S.}~\bibnamefont {Maekawa}},\ }\href {\doibase
  10.1103/PhysRevB.87.014423} {\bibfield  {journal} {\bibinfo  {journal}
  {Physical Review B}\ }\textbf {\bibinfo {volume} {87}},\ \bibinfo {pages}
  {014423} (\bibinfo {year} {2013})}\BibitemShut {NoStop}%
\bibitem [{\citenamefont {Pe{\c c}anha-Antonio}\ \emph
  {et~al.}(2022)\citenamefont {Pe{\c c}anha-Antonio}, \citenamefont
  {Prabhakaran}, \citenamefont {Balz}, \citenamefont {Krajewska},\ and\
  \citenamefont {Boothroyd}}]{Pecanha-Antonio:2022}%
  \BibitemOpen
  \bibfield  {author} {\bibinfo {author} {\bibfnamefont {V.}~\bibnamefont
  {Pe{\c c}anha-Antonio}}, \bibinfo {author} {\bibfnamefont {D.}~\bibnamefont
  {Prabhakaran}}, \bibinfo {author} {\bibfnamefont {C.}~\bibnamefont {Balz}},
  \bibinfo {author} {\bibfnamefont {A.}~\bibnamefont {Krajewska}}, \ and\
  \bibinfo {author} {\bibfnamefont {A.~T.}\ \bibnamefont {Boothroyd}},\ }\href
  {\doibase 10.1103/PhysRevB.105.104422} {\bibfield  {journal} {\bibinfo
  {journal} {Physical Review B}\ }\textbf {\bibinfo {volume} {105}},\ \bibinfo
  {pages} {104422} (\bibinfo {year} {2022})}\BibitemShut {NoStop}%
\bibitem [{\citenamefont {Wickersheim}\ and\ \citenamefont
  {White}(1960)}]{Wickersheim:1960}%
  \BibitemOpen
  \bibfield  {author} {\bibinfo {author} {\bibfnamefont {K.~A.}\ \bibnamefont
  {Wickersheim}}\ and\ \bibinfo {author} {\bibfnamefont {R.~L.}\ \bibnamefont
  {White}},\ }\href {\doibase 10.1103/PhysRevLett.4.123} {\bibfield  {journal}
  {\bibinfo  {journal} {Physical Review Letters}\ }\textbf {\bibinfo {volume}
  {4}},\ \bibinfo {pages} {123} (\bibinfo {year} {1960})}\BibitemShut {NoStop}%
\bibitem [{\citenamefont {Wickersheim}(1961)}]{Wickersheim:1961}%
  \BibitemOpen
  \bibfield  {author} {\bibinfo {author} {\bibfnamefont {K.~A.}\ \bibnamefont
  {Wickersheim}},\ }\href {\doibase 10.1103/PhysRev.122.1376} {\bibfield
  {journal} {\bibinfo  {journal} {Physical Review}\ }\textbf {\bibinfo {volume}
  {122}},\ \bibinfo {pages} {1376} (\bibinfo {year} {1961})}\BibitemShut
  {NoStop}%
\bibitem [{\citenamefont {Lahoubi}\ and\ \citenamefont
  {Ouladdiaf}(2015)}]{Lahoubi:2015}%
  \BibitemOpen
  \bibfield  {author} {\bibinfo {author} {\bibfnamefont {M.}~\bibnamefont
  {Lahoubi}}\ and\ \bibinfo {author} {\bibfnamefont {B.}~\bibnamefont
  {Ouladdiaf}},\ }\href {\doibase https://doi.org/10.1016/j.jmmm.2014.02.015}
  {\bibfield  {journal} {\bibinfo  {journal} {Journal of Magnetism and Magnetic
  Materials}\ }\textbf {\bibinfo {volume} {373}},\ \bibinfo {pages} {108}
  (\bibinfo {year} {2015})}\BibitemShut {NoStop}%
\bibitem [{\citenamefont {Lahoubi}(2012{\natexlab{b}})}]{Lahoubi:2012b}%
  \BibitemOpen
  \bibfield  {author} {\bibinfo {author} {\bibfnamefont {M.}~\bibnamefont
  {Lahoubi}},\ }in\ \href {\doibase 10.5772/36317} {\emph {\bibinfo {booktitle}
  {Neutron Diffraction}}},\ \bibinfo {editor} {edited by\ \bibinfo {editor}
  {\bibfnamefont {I.}~\bibnamefont {Khidirov}}}\ (\bibinfo  {publisher}
  {IntechOpen},\ \bibinfo {address} {Rijeka},\ \bibinfo {year} {2012})\
  Chap.~\bibinfo {chapter} {10}\BibitemShut {NoStop}%
\bibitem [{\citenamefont {Sayetat}\ \emph {et~al.}(1984)\citenamefont
  {Sayetat}, \citenamefont {Boucherle},\ and\ \citenamefont
  {Tcheou}}]{Sayetat:1984}%
  \BibitemOpen
  \bibfield  {author} {\bibinfo {author} {\bibfnamefont {F.}~\bibnamefont
  {Sayetat}}, \bibinfo {author} {\bibfnamefont {J.}~\bibnamefont {Boucherle}},
  \ and\ \bibinfo {author} {\bibfnamefont {F.}~\bibnamefont {Tcheou}},\ }\href
  {\doibase https://doi.org/10.1016/0304-8853(84)90360-3} {\bibfield  {journal}
  {\bibinfo  {journal} {J. Magn. Magn. Mater.}\ }\textbf {\bibinfo {volume}
  {46}},\ \bibinfo {pages} {219} (\bibinfo {year} {1984})}\BibitemShut
  {NoStop}%
\end{thebibliography}

%

\end{document}